# Bibliometric Patterns and Concept Evolution Trajectories in research publications in 'Future Generation Computer Systems'

Prashasti Singh[a], Vivek Kumar Singh[a,1], Hiran H. Lathabai[b]

[a]Department of Computer Science, Banaras Hindu University, Varanasi, Uttar Pradesh, India – 221005.
[b]DST-Centre for Policy Research, Indian Institute of Science, Bengaluru, Karnataka, India – 560012.

**Abstract:** Future Generation Computer Systems (FGCS), published by Elsevier, is a prestigious peer reviewed journal that started in 1984. As on date of writing this article, the journal is in its 137[th] volume. Owing to its publication quality and continued academic standards, it has been indexed by major academic databases such as Web of Science, Scopus, DOAJ etc. It is now ranked among the top journals in the area of Computer Science (General). Motivated by the long history, academic reputation and prestige of the journal, the present study attempts to do a bibliometric and text-based analysis of publications in the journal during the period of 1984-2020 (37 years). Bibliometric analysis helped to identify publishing and citation patterns, authorship and collaboration structure, funding patterns of the published research, open access and altmetric impact, gender distribution and SDG connections etc. The text-integrated path analysis helped in identifying the thematic structure and major concept evolution trajectories. The results of the analysis indicate that the FGCS journal is a high-quality research journal that has grown over time in terms of publications, citations and rankings. It publishes articles in specified thematic areas, with cloud, grid, IoT, Blockchain etc. found as the major themes addressed. Knowledge extension from cloud computing to multi-domain networks is witnessed and a divergence towards 'service function chains', 'distributed machine learning' and 'edge computing systems' is visible. The article thus presents very informative and useful insight about research publication trends in FGCS.

## 1. Introduction

Future Generation Computer Systems, published by Elsevier, is a prestigious peer reviewed journal that started in 1984. As on date of writing this article, the journal is in its 137[th] volume. According to Aims & Scope document of the journal, it mainly publishes articles in the areas of distributed systems, collaborative environments, high performance and high-performance computing, Big Data on such infrastructures as grids, clouds and the Internet of Things (IoT). It mainly publishes full length articles, survey/ review articles and extended versions of conference/ workshop articles. It is a hybrid journal that supports an APC based open access publication model.

Owing to its publication quality and continued academic standards, it has been indexed by major academic databases such as Web of Science, Scopus, DOAJ etc. For example, it had an Impact factor of 1.864 in 2012, which has increased to 7.307 in 2021, a growth of almost 4 times in 10 years. Its 5-year impact factor is 6.95. As per the latest Journal Citation Report, it features in Q1 in Computer Science-Theory and Methods Category and is ranked 10/109.

[1] Corresponding author. Email: vivek@bhu.ac.in

According to Scopus, it is ranked in the 97[th], 98[th] and 99[th] percentile in Computer Science-Software (10[th] rank out of 398), Computer Science- Computer Networks & Communication (5[th] out of 359) and Computer Science- Hardware & Architecture (2[nd] out of 167), respectively. It has a cite score of 18.7 and a Scimago Journal Rank (SJR) of 2.233 (Q1) as in 2021. Its Source Normalized Impact per Paper (SNIP) is 2.699 in 2021. According to the latest Google Scholar Metrics of 2022, it is ranked 2[nd] in Computing Systems, a subcategory of Engineering and Computer Science journals, with an h5 index of 133 and h5 median of 197.

Motivated by the long history, academic reputation and prestige of the journal, the present study attempts to do a bibliometric and text-based analysis of publications in the journal during the period of 1984-2020 (37 years). To the best of our knowledge, no previous study has tried to analyse the publication patterns, authorship and citation trends, and thematic trajectories of publications in the journal. Therefore, this study attempts to do a systematic bibliometric analysis of the publications in the journal to understand the publication, citation, authorship, gender distribution, open access and social media visibility trends. In addition, it dives deep into the major evolutionary trajectories of the thematic focus of the publications in the journal, as captured by the 'Text-integrated Path Analysis (TPA)' method. The analysis is thus divided into two parts. The first part presents useful insights about bibliometric patterns and trends. The second part uncovers the thematic trajectories seen in the publications in the journal.

The bibliometric exercise presents useful analysis and insights about the following major aspects:

- Publication volume and annual trends and growth during 1984-2020,
- The citation impact of publications in the journal,
- Top contributing countries and institutions that contributed publications in FGCS,
- Authorship structure and international collaboration patterns in publications in FGCS,
- Gender distribution of publications in FGCS,
- Funded and Non-funded research published in FGCS,
- Open access availability of publications in FGCS,
- Social media visibility of publications in FGCS on different platforms, and
- Connect of FGCS publications with the Sustainable Development Goals.

The Text-integrated Path Analysis performs a network and text-based analysis and provides insights about following major things:

- Key developmental trajectories within the body of knowledge represented by FGCS,
- Evolution of key concepts related to the fields that are associated with the journal FGCS,
- Some key application areas in which the specific developments (that are identified based on the network mining exercise through path analysis) disseminated through FGCS can make a difference,
- Technological evolution in those key application areas associated with the body of knowledge represented by FGCS,
- Knowledge and technological evolution through convergences and divergences between the core areas and application areas, and
- Technologies that can bring paradigm shift and revolutionize immediate and indirect/more generic application areas of FGCS

The rest of the article is organized as follows: **Section 2** presents the Data and Methodology used for the analytical exercise. **Section 3** presents results about different aspects of bibliometric analysis. **Section 4** presents results of Text-integrated Path Analysis. The paper concludes in **Section 5** with a summary of the work done and the key conclusions obtained.

## 2. Data and Method

The data for analysis corresponds to publications in the FGCS journal during 1984 to 2020 and was obtained from the Dimensions database through a subscription-based api access. Dimensions database is known to have many advantages for bibliometric analysis and has been found useful for bibliometric analysis, in various respects [1], [2]. A total of 5,754 publication records were retrieved. The complete metadata comprising the publication records was downloaded and processed. The major fields used for analysis were 'doi', 'year', 'id', 'authors_count', 'authors', 'research_orgs', 'research_org_countries', 'referenced_pubs', 'open_access', 'category_for', 'category_sdg', 'concepts_scores', 'times_cited', 'funders' etc. Out of these, the 'open_access' field provides evidence about open access availability of the article. The 'category_for' field provides the subject classification of the publication according to a predefined scheme. The 'concept_scores' field lists important concepts (with relevance score) of a publication extracted from its full-text using machine learning algorithms. Other metadata fields are the standard fields used in a bibliometric analysis.

The data for social media activity around publications in FGCS was obtained from popular Altmetric aggregator Altmetric.com. The subscription-based access of Altmetric.com API was used for the purpose. It crawls over different social media, news, and blog platforms to collect views, mentions, comments etc. around a scholarly article. The Altmetric.com provides the mentions of publication DOIs on various social media platforms such as Twitter, Facebook, Wikipedia, Blogs etc. The DOIs of the publication data in FGCS journal were passed to Altmetric Explorer and the data corresponding to publications was obtained. Out of 5,754 publication records that were passed to Altmetric.com, it returned data for 1,710 publications, indicating that altmetric data is available only for publications in the recent decade.

### *2.1 Bibliometric, Altmetric and Gender Distribution Analysis*

The present study uses a combination of standard bibliometric approach, network theoretic approach and text analysis-based approach for mapping the bibliometric patterns and thematic structure of the publications in FGCS. A bibliometric study of a journal is a popular approach for identifying the trends of the journal in terms of topics, contributing institutions, authorship structure and highly cited papers etc. Several previous studies have adopted a bibliometric analysis approach for understanding publication patterns and thematic structure of a journal [3], [4], [5], [6].

For performing bibliometric analysis, a set of standard procedures were implemented in Python programming language. Basic indicators and results such as publication counts, citation impact, authorship structure, collaboration trends, funding patterns, etc were computed. Additionally, results such as top contributing countries and institutions, open access visibility and publications classified under Sustainable Development Goals were also computed.

The altmetric analysis for publications from FGCS was carried out by analysing the data obtained from Altmetric.com. This service provides 46 fields in the data, including DOI, Title, Twitter mentions, Facebook mentions, News mentions, Altmetric Attention Score, OA Status,

Subjects (FoR), Publication Date, URI, etc. Out of this, mainly data for Twitter, Facebook, News, Blog, Wikipedia and Mendeley platforms was analysed. The coverage percentage and average mentions per paper was identified.

In addition to standard bibliometric and altmetric patterns, the gender distribution in publications of FGCS was also identified. For this purpose, the service of gender-api was used. The name of the first author of each publication along with affiliating country name was extracted from publication data and passed to gender-api. The service returns the identified gender of the author with an accuracy score. For this study, an accuracy value above 70% has been considered acceptable. The obtained gender information for the publications was used to plot the year-wise gender distribution of authors.

A diagrammatic representation of the complete methodology for analysis, comprising of Bibliometric, Altmetric, Gender Distribution and Path Analysis steps of the FGCS publications is shown in **Figure 1**.

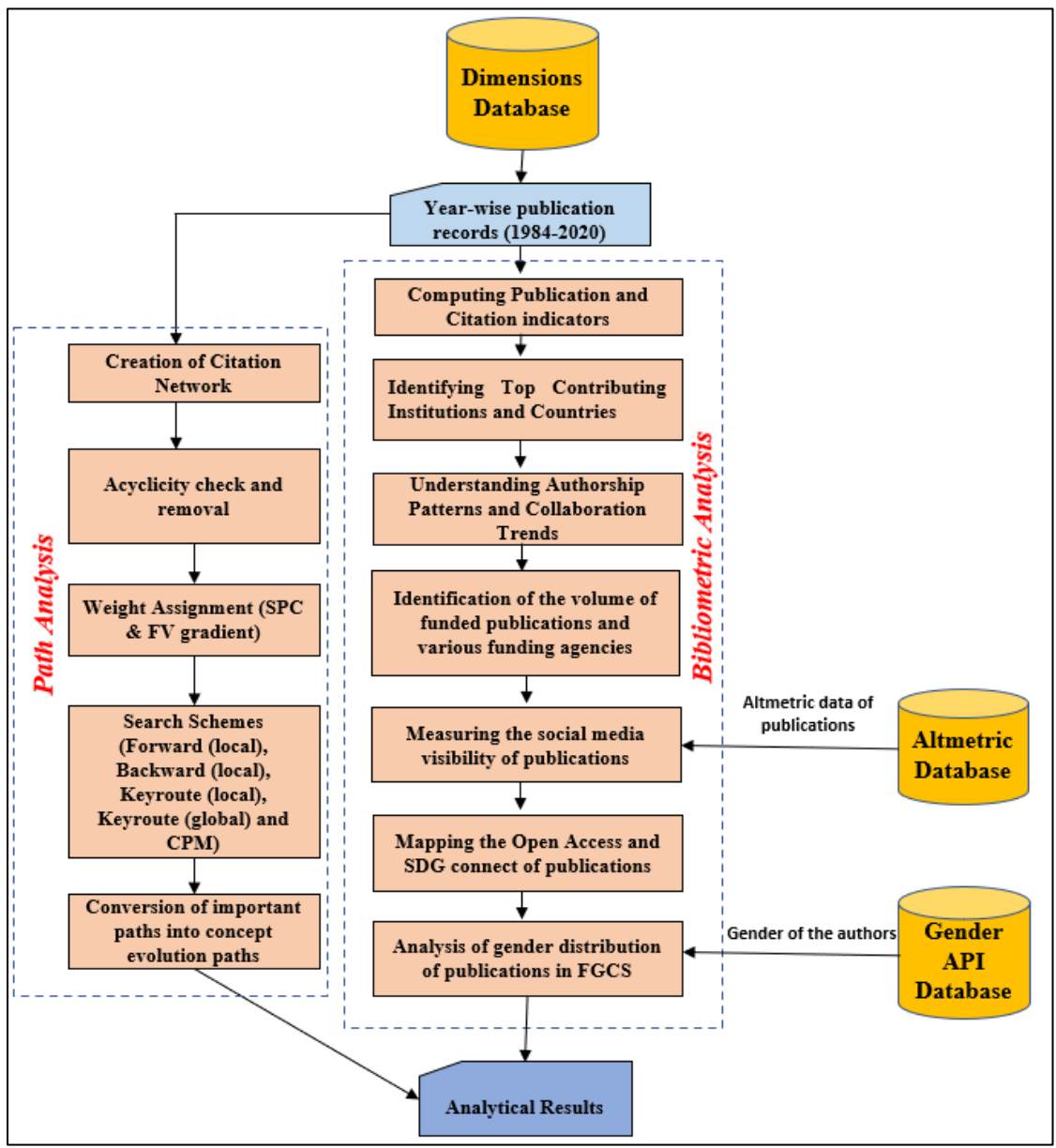

**Fig. 1: Methodological Approach**

## 2.2 Text-integrated Path Analysis (TPA)

Text-integrated path analysis is a methodology developed by combined usage of path analysis and text analysis. Firstly, the path analysis methodology is discussed in detail.

*Path Analysis*

Complex network analysis offers a multitude of methods and tools for addressing wide range of problems in real-world systems that can be modelled as networks. Networks are basically graphs (that consists of vertices that represent entities and their relationships represented as interconnections/links) and information/attributes related to both the vertices and links. Scientific and technological (patent) literature is two such systems that consists of several relationship between entities such as works (papers or patents), authors/inventors, institutional affiliations/assignees, etc. One of the major relationships is citations, where for a pair of works, the latest work cites previous related work. This results in the formation of scientific citations network or citation network of scholarly publications/papers [7], [8] and patent citation networks or citation networks of patents. Centrality analysis, path analysis and cluster analysis are three major kinds of analyses possible in network analysis that can be dubbed as the network analysis trio. Among these, path analysis helps to identify the important evolutionary trajectories or connective threads of knowledge [9]. Path analysis has been used by many for determining important evolutionary paths that might have served as backbone of development of different fields. Few examples are evolutionary trajectory study for medical technologies [10], 'Coronary angioplasty' [11], human resource development [12] and archaeology [13], etc. Despite its benefits, when it comes to methodical development, path analysis remained underexplored among the network analysis trio.

Citation networks are by default unweighted. Path analysis consists of two major steps- (i) weight assignment or conversion of unweighted network into weighted network and (ii) search or trace through the weighted network to retrieve paths. Original Hummon & Dorain framework [9] used traversal as the basis of weight assignment of arcs (directed citation links) for the creation of weighted network. They introduced SPLC (Search Path Link Count) and SPNP (Search Path Node Pair) as two weight assignment methods. Regarding search schemes, Forward Search and Critical path method were used for retrieving important evolutionary paths in the network. After this, a break-through in path analysis literature was brought-in by [14] when computationally efficient traversal-based weight assignment method namely SPC (Search Path Count) method was introduced. Generally, the traversal-based weight assignment schemes can be termed together as SPX schemes. A major development in search scheme happened with the introduction of search schemes like backward search, key-route search, global search, etc., with a provision for tolerance. These schemes, especially multiple key-route search, empowered the ability of path analysis to retrieve more important evolutionary trajectories in scientific literature. However, due to the dependency on SPX methods on the global structure of the network, in a connected component, highly (SPX) weighted arcs seem to clutter together, thereby limiting the effectiveness of multiple keyroute search to retrieve multiple paths. In order to address this limitation, [15] introduced a novel weight assignment method, namely the Flow Vergence (FV) gradient method for weight assignment and also upgrading the Liu-Lu approach [16]. with state-of-the-art integrated framework in path analysis. With the framework of [15], if we use SPX and FV gradient as weight assignment methods, and existing search schemes, following paths can be retrieved: SPX-forward paths, SPX-backward paths, SPX-keyroute paths, SPX-critical paths, FV-forward paths, FV-backward paths, FV-keyroute paths and FV-critical paths. In this work, we use SPC (among SPX) weight assignment and therefore the retrieved paths will be SPC-forward path, SPC-backward path, SPC-keyroute path, SPC-critical path, FV-forward path, FV-backward path, FV-keyroute path and FV-critical path.

Brief description of SPC and FV gradient weight assignment and search schemes are given below:

*SPC method: A short revisit*

As mentioned earlier, SPC method is one of the weight assignment methods that belong to the SPX genre (traversal-based weight assignment methods) developed by [14] In SPC method, identification of all sources and sinks in the citation network is the first task. Then, for each arc in the network, the number of search paths (from sources to sinks) passing through it is computed and such a count is assigned as weight of that arc.

*FV gradient method: A short revisit*

Scientific citation network, a kind of information network is an interconnected structure of scholarly publications which are linked through citation links or arc of citations. These links can be treated as a representation of the flow of information from the cited work to the citing work. Most of the vertex measures if used independently fails to reflect one important property of a vertex (work) in a citation network that arises due to the flow of information through it, namely the flow vergence (FV). Flow of information through a paper invokes a dominance in terms of flow convergence or flow divergence, which can be termed as flow vergence by [17]. Flow vergence potential is the potential of a paper to contribute to the growth of a field via the attainment or improvement of flow divergence dominance. Thus, works which are presently in knowledge flow divergence mode can be surely regarded as works of high flow vergence potential. Even though a work is not presently in flow divergence mode, certain works can be treated as of high FV potential. This happens only if that work succeeded in delivering knowledge to other works (that cites the concerned work) that are at certain level of quality which is reflected by eigenvector centrality ([18]). This concept was introduced by [17] and an index for reflecting the flow vergence potential was also developed. Flow Vergence index or FV index of a paper *i* can be computed as:

$$W_{FV_i} = \frac{indeg_i - outdeg_i}{indeg_i + outdeg_i} + eig_i \qquad (1)$$

where, $indeg_i$, $outdeg_i$ and $eig_i$ are the indegree (the number of citations received), the outdegree (number of citations made) and eigenvector centrality of paper *i* in the network [19]. Now, we move on to discuss the FV gradient method as the theoretical origin and rationale behind the formulation of the FV index is already covered by [15], [17], [20], [21], and [22].

Since a citing-cited pair is related by knowledge flow, existence of flow vergence potential for citing and cited work also implies the existence of flow vergence gradient or gradient in flow vergence potential between them. The potential difference between works *i* and *j* connected by an arc of citation from *j* to *i* can be termed as FV gradient ([20]) and it can be computed as:

$$\Delta FV_{ij} = W_{FV_i} - W_{FV_j} \qquad (2)$$

The speciality of the of FV gradient or the FV potential difference is that, usually, the knowledge flow tends to appear as to have occurred from a work of high FV potential to a work of low FV potential. Thus, in most of the cases, $\Delta FV_{ij}$ takes a positive value. However, in some special cases, when newer works (citing works) tend to outperform the former ones (cited ones) due to its intellectual merit or innovative appeal to the scientific community, we can have $\Delta FV_{ij} < 0$, making knowledge flow to occur from low potential paper to a high potential paper. This phenomenon was termed as Flow Vergence effect or FV effect and was used to detect [20] and predict [22] pivot papers of paradigm shift. Computation of FV gradients makes the network a signed weighted network and hence FV gradient method of weight assignment helps for the retrieval of paths that might not be highlighted through SPX methods. To ensure that the arcs with negative signs are not missed and given high priority when certain search schemes like key-route search are employed, the following transformation of eqn. (2) should be done.

$$\Delta FV_{ij}(norm.) = 1 + \frac{max.\Delta FV - \Delta FV_{ij}}{max.\Delta FV - min.\Delta FV} \quad (3)$$

Where, $max.\Delta FV$ is the highest FV gradient weight in the network and $min.\Delta FV$ is the lowest FV gradient weight in the network.

*Search Schemes*

**Forward Search**: In forward search, among all the arcs originating from all the sources (the papers that do not cite any other paper but get at least one citation), the ones with highest weight will be selected. The target node of that arc will be made the new source and the same procedure will be repeated in a greedy fashion till sink papers (the papers that do not get cited yet but has cited at least one paper) are found. When there are ties, both the arcs will be considered.

**Backward Search**: In backward search, unlike the forward search, search originates from sink papers till source papers are obtained. Everything is same as that of forward search. This was introduced by [16].

**Global search/Critical path method**: In global search method, instead of selecting an initial arc and its subsequent arcs in a greedy or 'local best' fashion, for all the source-sink paths, the total weight of the path (which is equal to the sum of weights of all the arcs that forms that path) will be computed and the highest among that will be considered as the critical path.

**Key-route Search (Local)**: In key-route search (local), instead of initiating search from source or sink (which invokes the risk of missing the highest weighted arc in the network), search commences from the terminal nodes of key-route (the highest weighted arc). From cited work of the key-route, search proceeds as that in backward search till source is obtained and from citing work of the key-route, search proceeds in the manner of forward search till sink is found. This search scheme is the key innovation of Liu-Lu approach. This search can be conducted to retrieve multiple paths with the selection of multiple key-routes. For instance, the software package PAJEK () has set the option 1-10 as default choice, by which top 10 key-routes will be selected for the search. This choice can be changed by users accordingly.

**Key-route Search (global)**: In key-route (global), instead of tracing from both the terminal nodes of key-routes till sources and sinks are approached, a global search is initiated from terminal nodes of key-routes. That means, from all the paths reachable from citing paper and cited paper in the key-route, the ones with largest sum of weights will be chosen.

Along with these search schemes, a provision for tolerance is also proposed by [16] and available in PAJEK for local search schemes such as forward, backward and key-route (local). If a tolerance of 10% (0.1 is the default value set in PAJEK) is selected, at each step, instead of choosing the largest weighted arc, all the arcs that falls with-in 90% of its weight will be selected.

*Text-integration for Path Analysis*

Once, important paths (which are subnetworks of citation network) in the concerned literature are identified, nodes of the paths (i.e., papers) should be relabelled or current vertex labels should be replaced by terms or words that best represent the theme/contribution of that paper.

In Dimensions, since each published article is associated with 'concepts' that represent certain keywords that represent the relation of work with the subfield/field in which the work belongs to, the relabelling can be easily achieved by mapping the work's label/id with the concept. Also, since each concept are associated with a relevance score or score of relevance, selection of the best representative word can also be systematically done with ease. Thus, the paths obtained after TPA can be termed as 'Concept Evolution Paths'. The methodology used for text-integrated path analysis (TPA) is given below:

**Procedure for extraction of Concept evolution Paths**

**Input**: Citation network of published articles $C_i$ and Work-Concept/Keyword affiliations network $WK$.

**Output**: Concept evolution paths

1. From $C_i$, extract SPC paths and FV paths
2. For every pair of paths $P_i$ and $P_j$

   I. Compute $U_{P_i P_j} = \frac{|P_i|+|P_j|-|P_i \cap P_j|}{|P_i|+|P_j|}$, the uniqueness index ([15]) of $P_i$ with respect to $P_j$ and vice-versa.

   II. a) If $U_{P_i P_j} \geq \delta$ (desirable value is 0.65), select both

   b) Otherwise,
   
   (i) Select path $P_i$ if $|P_i| > |P_j|$
   
   (ii) Select path $P_j$ if $|P_j| > |P_i|$
   
   (iii) Select both if $|P_i| = |P_j|$

3. From $W - K$ network, extract the subnetwork $W - K(P_x)$ (where $P_x$ is the selected path at step 3) network by choosing concept with highest relevance score
4. Obtain concept evolution paths or concept citation path $K(P_x) \to K(P_x)$ using

$$K(P_x) \to K(P_x) = (W - K(P_x))^T \times P_x \times (W - K(P_x))$$

## 3. Bibliometric Patterns

*3.1 Publications and Citations*

The FGCS journal published a total of 5,754 research items during the 1984-2020 period, all of which taken together got 128,337 citations till date. **Table 1** presents the various publication and citation indicators for publication records in FGCS. Publication indicators such as year-wise TP (Total Publications) and growth rate of publications for the 37-year time span namely AGR (Annual Growth Rate) and CAGR (Compounded Annual Growth Rate) are computed and reported. The Compounded Annual Growth Rate (CAGR) is calculated as follows:

$$\text{CAGR} = \left( \left( \frac{Vfinal}{Vbegin} \right)^{\frac{1}{t}} - 1 \right) * 100$$

where, V*final* is the number of publication records in the year 2020, V*begin* is the number of publication records in the year 1984, and t is the time period in years.

**Table 1: Publication and Citation trends for FGCS**

| Year | TP | AGR (%) | CAGR (%) | TC | CPP | Cited (%) | h-index |
|------|-----|---------|----------|------|------|-----------|---------|
| 1984 | 18 | -- | | 56 | 3.11 | 55.56 | |
| 1985 | 23 | 27.78 | | 143 | 6.22 | 60.87 | |
| 1986 | 37 | 60.87 | | 85 | 2.30 | 48.65 | |
| 1987 | 40 | 8.11 | | 136 | 3.40 | 52.5 | |
| 1988 | 19 | -52.5 | | 256 | 13.47 | 94.74 | |
| 1989 | 48 | 152.63 | | 149 | 3.10 | 66.67 | |
| 1990 | 36 | -25 | | 106 | 2.94 | 61.11 | |
| 1991 | 18 | -50 | | 33 | 1.83 | 72.22 | |
| 1992 | 71 | 294.44 | | 197 | 2.77 | 45.07 | |
| 1993 | 37 | -47.89 | | 49 | 1.32 | 45.95 | |
| 1994 | 52 | 40.54 | | 150 | 2.88 | 53.85 | |
| 1995 | 59 | 13.46 | | 260 | 4.41 | 52.54 | |
| 1996 | 27 | -54.24 | | 387 | 14.33 | 81.48 | |
| 1997 | 39 | 44.44 | | 1333 | 34.18 | 71.79 | |
| 1998 | 68 | 74.36 | | 613 | 9.01 | 73.53 | |
| 1999 | 94 | 38.24 | | 2400 | 25.53 | 87.23 | |
| 2000 | 85 | -9.57 | | 3848 | 45.27 | 81.18 | |
| 2001 | 97 | 14.12 | | 1505 | 15.52 | 83.51 | |
| 2002 | 73 | -24.74 | 10.87 | 1378 | 18.88 | 87.67 | 142 |
| 2003 | 136 | 86.3 | | 1397 | 10.27 | 78.68 | |
| 2004 | 122 | -10.29 | | 2818 | 23.10 | 87.7 | |
| 2005 | 140 | 14.75 | | 2569 | 18.35 | 85.71 | |
| 2006 | 108 | -22.86 | | 1706 | 15.80 | 85.19 | |
| 2007 | 107 | -0.93 | | 2344 | 21.91 | 91.59 | |
| 2008 | 83 | -22.43 | | 1542 | 18.58 | 93.98 | |
| 2009 | 105 | 26.51 | | 6397 | 60.92 | 98.1 | |
| 2010 | 150 | 42.86 | | 3076 | 20.51 | 92.67 | |
| 2011 | 124 | -17.33 | | 3078 | 24.82 | 94.35 | |
| 2012 | 134 | 8.06 | | 7237 | 54.01 | 93.28 | |
| 2013 | 199 | 48.51 | | 13403 | 67.35 | 96.48 | |
| 2014 | 237 | 19.1 | | 6237 | 26.32 | 97.47 | |
| 2015 | 134 | -43.46 | | 4081 | 30.46 | 100 | |
| 2016 | 260 | 94.03 | | 7519 | 28.92 | 98.08 | |
| 2017 | 314 | 20.77 | | 7985 | 25.43 | 97.77 | |
| 2018 | 792 | 152.23 | | 22046 | 27.84 | 98.11 | |
| 2019 | 848 | 7.07 | | 14653 | 17.28 | 95.17 | |
| 2020 | 820 | -3.3 | | 7165 | 8.74 | 88.41 | |

Notes: TP - Total Publications in a Year; TC- Total Citations; CPP – TP/TC; Cited %- Publications having >0 citations

It is observed that during the given period (1984 to 2020), the number of publications grew by a factor of 45, starting from 18 publications in 1984 and reaching to 820 publications in 2020. The growth in number of publications during the first decade was, however, slower as compared to later periods. A steady increase in publication counts was noticed for the recent decade (growing from 150 publications in 2010 to 820 publications in 2020). The AGR is found to vary from 27.78% during initial period to 152.23% in 2017-2018. High AGRs were observed during 1988-1989 (152.63%), 1991-1992 (294.44%) and 2017-2018 (152.23%). However, the number of publications also experienced a decline in few time periods. For example, a decrease of more than 40% in publication count was observed during the time periods of 1987-1988 (-52.5%), 1990-1991 (-50%), 1992-1993 (-47.89%), 1995-1996 (-54.24%) and 2014-2015 (-43.46%). However, in overall terms, the publications in FGCS have grown with an overall CAGR of 10.87%.

Next, the citation patterns of publications in FGCS are analysed. In the scientific community, citations are an important indicator of the impact of publications produced by researchers, institutions as well as countries [23]. They account for the scientific impact of a journal in which the publications are published. An overview of citation impact of publications in FGCS can be seen in **Table 1.** Various indicators, namely TC (Total Citations), CPP (Citation per paper given by TC/TP), Cited% (percentage of publications having at-least 1 citation) and h-index of the journal are computed and shown. It can be observed that during the initial phase of the journal, a steady increase in total citation count was recorded, growing from 56 in 1984 to 256 in 1988. The CPP grew from 3.11 in 1984 to 13.47 in 1988. In the subsequent years, very low CPP values were observed, which ultimately spiked to 34.18 in 1997. Since it is a ratio of TC to TP, it will tend to decrease with fewer citations and more publications. Similarly, it will tend to increase with more citations and fewer publications, thus directly implicating the high quality and importance of these few publications. When TC and TP go up or down together, it does not affect much change in CPP. CPP value of greater than 50 was observed for the years 2009 (60.92), 2012 (54.01) and 2013 (67.35). In these years, the publications were relatively less but they obtained a good number of citations. The growing popularity of the journal could also be noticed the fact that since 1996 more than 70% of the articles received at least one or more citations. The h-index of the journal computed for the whole period is 142, indicating that there are at least 142 papers, each of which attracted 142 or more citations.

*3.2 Top Contributing Countries and Institutions*

The bibliometric analysis is further continued to identify the major countries and institutions that contributed research publications to FGCS. **Table 2** presents the top ten contributing countries to the FGCS journal during 1984 to 2020. China accounts for majority of the publications (1,434) and citations (29,335) in the journal. United States is at $2^{nd}$ place in publications (1,068) and citations (26,944). Though China and United States accounted for the maximum number of publications and citations accrued, the highest CPP value of 60.95 is observed for Australia which had only 427 publications in FGCS. India which stood at $9^{th}$ position in publications with 286 publications accounted for the $2^{nd}$ highest CPP value of 33.02 after Australia.

**Table 3** shows the list of top ten contributing institutions to FGCS journal. Out of these ten institutions, four belong to China. There are two institution each from Australia and Italy each. Out of the top ten institutions, the Huazhong University of Science and Technology (China), King Saud University (Saudi Arabia) and University of Amsterdam (Netherlands) have each contributed more than 100 papers. The Huazhong University of Science and Technology (China) is at the $1^{st}$ position among contributing institutions with 122 papers in FGCS. It can be observed that the University of Melbourne (Australia) has the highest number of citations accrued, with a very good CPP value of 227.24. University of Naples Federico II from Italy gathered the $2^{nd}$ highest number of citations with just 52 publications.

**Table 2: Top contributing Countries**

| Country | TP | TC | CPP | Cited % |
|---|---|---|---|---|
| China | 1434 | 29335 | 20.46 | 94.07 |
| United States | 1068 | 26944 | 25.23 | 92.42 |
| United Kingdom | 597 | 12948 | 21.69 | 89.45 |
| Italy | 442 | 8798 | 19.90 | 92.08 |
| Spain | 428 | 8733 | 20.40 | 94.63 |
| Australia | 427 | 26025 | 60.95 | 95.78 |
| France | 337 | 5228 | 15.51 | 85.76 |
| Germany | 297 | 5624 | 18.94 | 90.57 |
| India | 286 | 9443 | 33.02 | 96.15 |
| South Korea | 249 | 5289 | 21.24 | 91.57 |

Notes: TP- Total Publications; TC - Total Citations; CPP – TC/TP; Cited % – Publications having >0 citations

**Table 3: Top contributing Institutions**

| Institution | Country | TP | TC | CPP | Cited % |
|---|---|---|---|---|---|
| Huazhong University of Science and Technology | China | 122 | 1962 | 16.08 | 94.26 |
| King Saud University | Saudi Arabia | 107 | 2887 | 26.98 | 99.07 |
| University of Amsterdam | Netherlands | 106 | 1095 | 10.33 | 82.08 |
| Deakin University | Australia | 70 | 2950 | 42.14 | 94.29 |
| University of Melbourne | Australia | 62 | 14089 | 227.24 | 95.16 |
| Xidian University | China | 59 | 1057 | 17.92 | 93.22 |
| University of Calabria | Italy | 54 | 1681 | 31.13 | 98.15 |
| University of Electronic Science and Technology of China | China | 53 | 1596 | 30.11 | 92.45 |
| University of Naples Federico II | Italy | 52 | 1875 | 36.06 | 96.15 |
| Wuhan University | China | 47 | 699 | 14.87 | 100 |

Notes: TP- Total Publications; TC - Total Citations; CPP – TC/TP; Cited % – Publications having >0 citations

*3.3 Top Cited and Citing Sources*

The identification of top cited and top citing sources is the next step in bibliometric analysis. **Table 4** presents the list of top 25 cited sources (journals) and the top 25 citing sources (journals) for FGCS. It can be observed that the publications in FGCS frequently cite the articles published in Future Generation Computer Systems, Lecture Notes in Computer Science, IEEE Transactions on Parallel and Distributed Systems, IEEE Access, Communications of the ACM, Concurrency Computation Practice and Experience and IEEE Communications Magazine etc. Similarly, the publications in FGCS are cited quite frequently by the papers published in IEEE Access, Future Generation Computer Systems, Sensors, Concurrency and Computation Practice Experience, IEEE Internet of Things Journal, Journal of Supercomputing, Multimedia Tools and Applications, Cluster Computing etc. It can be seen that the citing sources are a bit more diverse than the cited sources in terms of their thematic focus. Thus, the publications in FGCS may be seen as contributing to research in other areas too. The list of top cited and citing sources thus provides a good indication of the trends of citation outflow and inflow for FGCS.

**Table 4: Top 25 Cited and Citing Sources for FGCS**

| Top Cited Journals (Cited by FGCS) | Citations | Top Citing Journals (Citing FGCS) | Citations |
| --- | --- | --- | --- |
| FUTURE GENERATION COMPUTER SYSTEMS | 9066 | IEEE ACCESS | 3840 |
| LECTURE NOTES IN COMPUTER SCIENCE | 5902 | FUTURE GENERATION COMPUTER SYSTEMS | 2849 |
| IEEE TRANSACTIONS ON PARALLEL AND DISTRIBUTED SYSTEMS | 1567 | SENSORS | 1519 |
| IEEE ACCESS | 1149 | CONCURRENCY AND COMPUTATION PRACTICE EXPERIENCE | 1068 |
| COMMUNICATIONS OF THE ACM | 1098 | IEEE INTERNET OF THINGS JOURNAL | 929 |
| CONCURRENCY AND COMPUTATION PRACTICE AND EXPERIENCE | 933 | JOURNAL OF SUPERCOMPUTING | 928 |
| IEEE COMMUNICATIONS MAGAZINE | 928 | MULTIMEDIA TOOLS AND APPLICATIONS | 705 |
| JOURNAL OF NETWORK AND COMPUTER APPLICATIONS | 872 | CLUSTER COMPUTING THE JOURNAL OF NETWORKS SOFTWARE TOOLS AND APPLICATIONS | 680 |
| IEEE TRANSACTIONS ON COMPUTERS | 824 | APPLIED SCIENCES BASEL | 663 |
| INFORMATION SCIENCES | 816 | EXPERT SYSTEMS WITH APPLICATIONS | 578 |
| COMPUTER NETWORKS | 802 | WIRELESS PERSONAL COMMUNICATIONS | 532 |
| JOURNAL OF PARALLEL AND DISTRIBUTED COMPUTING | 787 | LECTURE NOTES IN COMPUTER SCIENCE | 530 |
| EXPERTS SYSTEMS WITH APPLICATIONS | 735 | JOURNAL OF NETWORK AND COMPUTER APPLICATIONS | 523 |
| COMPUTER | 734 | JOURNAL OF AMBIENT INTELLIGENCE AND HUMANIZED COMPUTING | 507 |
| IEEE INTERNET OF THINGS JOURNAL | 731 | ELECTRONICS | 493 |
| IEEE COMMUNICATION SURVEYS & TUTORIALS | 698 | WIRELESS COMMUNICATIONS MOBILE COMPUTING | 484 |
| IEEE TRANSACTIONS ON KNOWLEDGE AND DATA ENGINEERING | 628 | SECURITY AND COMMUNICATION NETWORKS | 482 |
| PROCEEDINGS IEEE INFOCOM | 622 | INFORMATION SCIENCES | 452 |
| ACM COMPUTING SURVEYS | 594 | SUSTAINABILITY | 423 |
| IEEE TRANSACTIONS ON SERVICES COMPUTING | 557 | SOFT COMPUTING | 415 |
| JOURNAL OF GRID COMPUTING | 529 | APPLIED SOFT COMPUTING | 400 |
| ACM SIGCOMM COMPUTER COMMUNICATION REVIEW | 517 | COMPUTER NETWORKS | 394 |
| SENSORS-BASEL | 488 | COMPUTER COMMUNICATIONS | 385 |
| JOURNAL OF SUPERCOMPUTING | 483 | JOURNAL OF PARALLEL AND DISTRIBUTED COMPUTING | 369 |
| PROCEDIA COMPUTER SCIENCE | 478 | CMC COMPUTERS MATERIALS CONTINUA | 355 |

## 3.4 Authorship Structure

A global shift is seen from individualistic to team science [24], [25]. Further, the acceptance rate of publications involving collaborations among multiple authors has been found to be higher than single authored ones [26], [27]. Therefore, we tried to see the authorship structure of publications in FGCS over the years. **Figure 2** presents the year-wise authorship structure of publications in FGCS. The plot reveals that while single authored publications have been dominant in the early years of the journal (with more than 70% papers being single authored), the number of single authored papers has decline to even less than 10% in the recent years. The papers with 2-5 authors are now seen to be the most popular type, accounting for over 70% papers in the year 2020. Similarly, papers with 6-10 authors and 10+ authors being negligible till about 1990, show a rise in volume. The papers with 6-10 authors are about 20% in the recent years. Publications with 10+ authors, however, remain very small, with less than 5% of the total publications. The authorship structure observed shows a clear pattern of shift from single authored to multi-authored papers in the journal. Interestingly, the papers with 6-10 authors now constitute over 20% of the publication volume in the FGCS journal.

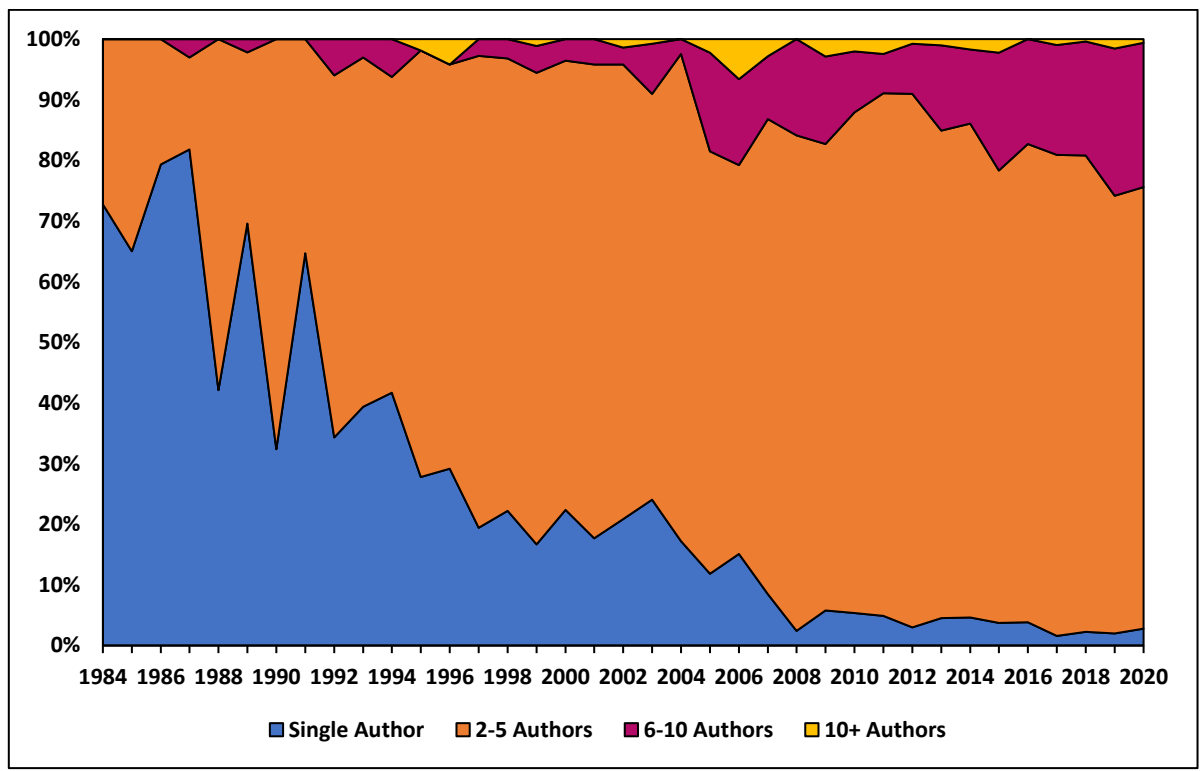

**Fig. 2: Authorship Structure (year-wise)**

## 3.5 Institutional and International Collaboration

It is well-known that research involving collaboration, especially international collaboration, benefits both the researchers and organizations and leads to research with higher impact. [28], [29]. Keeping this in mind, we have tried to find identify what proportion of FGCS multi-authored publications involve collaboration between authors from the same institution (Domestic - Single Institution), authors from different institutions from a single country (Domestic – Multi Institution) and authors from multiple institutions from different countries (Internationally Collaborated Papers (ICP)). The research_org_countries field in the publication data was analysed for this purpose. **Figure 3** presents the proportion of Domestic- Single Institution, Domestic- Multi Institution and ICP papers in FGCS over the period of time from 1984 to 2020. It can be observed that during the initial years, majority of the publications (about 80%) are Domestic- Single Institution type. However, after the year 2004, the ICP proportion crossed 20% mark and has increased continuously thereafter. During the years 2015-2020, ICP proportion has reached to over 40% level. It is in this same duration, that maximum number of citations were accrued by the journal's publications. This could be attributed to the increase in internationally collaborated papers during this time.

A look at the papers contributed as a part of domestic collaborations i.e. publications from single and multiple institutions reveals that the single institution papers occupied a high volume during the initial years of the journal, however their proportion declined after 2004. On the other hand, a rise in papers published as a result of collaboration among institutions from the same country increased from 1994 onwards. However, this increase remained moderate and gradual during the years. In the most recent years, about over 20 papers are examples of Domestic- Multi Institution type.

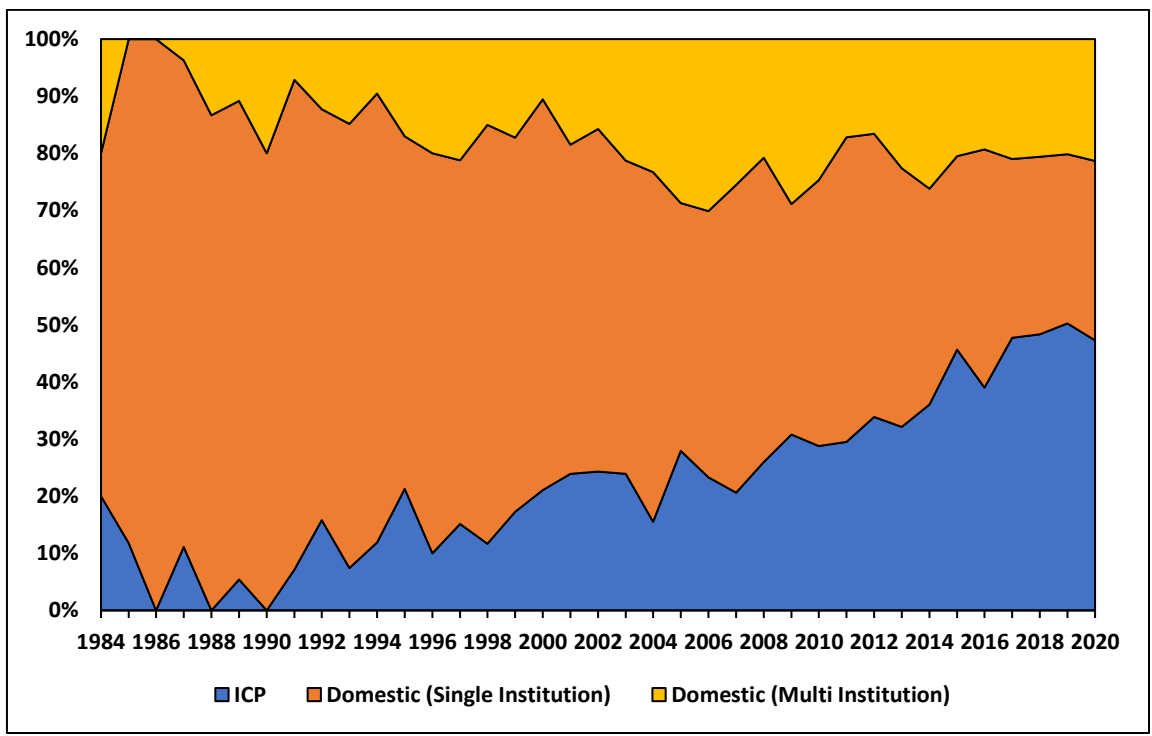

**Fig. 3 Collaboration Pattern (year-wise)**

*3.6 Funded Publications and Funding Agencies*

The publication metadata of publications in FGCS was analysed next to find out how many publications have associated funding, and which are the top funding agencies for such research. **Table 5** presents a list of major funding sources along with number of funded publications. It can be seen that the Chinese funding agency- National Natural Science Foundation of China- have highest number of funded publications in FGCS. Out of top ten funding agencies, three are from China. The remaining agencies in the list are from United States, Spain, South Korea etc. The European Commission, Belgium is the second largest contributor of funded publications.

**Table 5: Top Funding Sources**

| Funding Sponsor | Supported Publications |
|---|---|
| National Natural Science Foundation of China | 981 |
| European Commission, Belgium | 507 |
| Ministry of Science and Technology of the People's Republic of China | 343 |
| Directorate for Computer & Information Science & Engineering, United States | 202 |
| Ministry of Economy, Industry and Competitiveness, Spain | 195 |
| Ministry of Education of the People's Republic of China | 159 |
| National Research Foundation of Korea, South Korea | 94 |
| National Council for Scientific and Technological Development, Brazil | 90 |
| Engineering and Physical Sciences Research Council, United Kingdom | 86 |
| Ministry of Science and Technology, Taiwan | 70 |

*3.7 Open access availability*

The institutions across countries have been increasingly participating in open access publishing [30]. The availability of open access publications enhances greater participation in science for various audiences such as authors, researchers, funders etc. [31]. More so it is beneficial for developing and under-developed countries whose research fraternity may not have sufficient resources to pay for journal subscription charges [32]. There are different routes to open access, such as Gold, Green, Bronze etc. [33]. Owing to importance of open access availability of articles, several previous studies have tried to measure the extent and the type of open access publishing occurring at the level of a whole country too [34], [35]. Motivated by this, we have tried to analyzed the trend and pattern of open access availability of publications from FGCS. **Figure 4** presents year-wise trend of openly available artciles from FGCS. In the initial years, very few publications are available in open access. However, during the more recent time, about 20 to 30% articles are found to be available in some form of open access. The publications for the year 2016 show the peak of about 30% of the publications available in open access. **Figure 5** presents the distribution of openly available publications under different types. It can be observed that majority of the openly available publications (about 83%) are available in green form of open access. This is followed by 9% in bronze open access and 8% in hybrid open access form. Thus, there is an increase over the years in openly available articles from FGCS, which is now between 20 to 30% of the total articles published in FGCS.

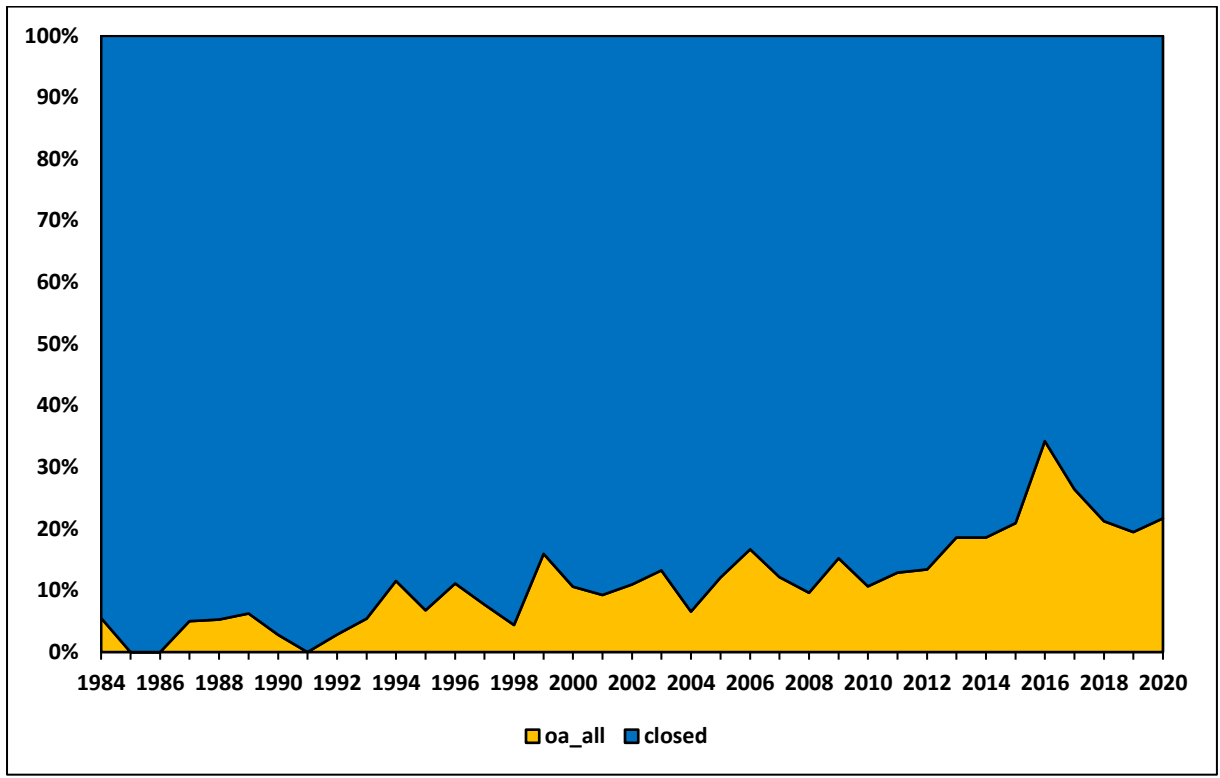

**Fig. 4: Open Access vs Closed Publications in FGCS during 1984-2020**

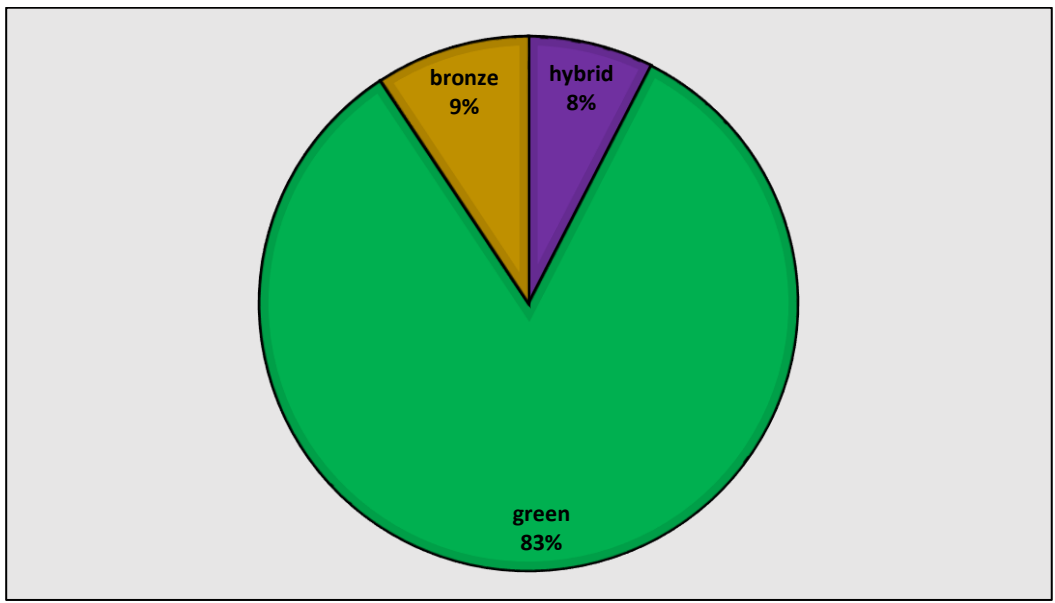

**Fig. 5: Open Access Types of open FGCS Publications**

*3.8 Gender distribution of research publications*

The gender distribution in research publishing is now a very important aspect in quantitative studies of science. Several studies have tried to determine the gender distribution in research either at the level of a country [36], [37] or a discipline [38]. Motivated by this, we have also tried to understand the gender distribution pattern of research publications from FGCS. The metadata for publication records, mainly the author names, have been analysed to find out what proportion of papers in FGCS are male first authored and what proportion of papers are female first authored. For this purpose, the service of gender-api was used, as described in the 'Data and method' section. The author names (along with country name) are passed to gender-api, which returns the determined gender for the author name with an accuracy score. The results for which a 70% accuracy value is obtained are accepted as valid determination of author gender. This information is then used to plot the year-wise gender distribution of papers, as seen in **Figure 6.** It can be seen that during the initial years, male first authored publications were high with almost negligible percentage of female first authored publications. However, the proportion of female first authored publication has rose after that, more specifically since the year 2000. While the female first authored publications were only 14.04% in 2000, it grew to a level of up to 20.09% in 2020. Thus, there is a growth seen in female first authored publications in FGCS, more specifically during the recent years.

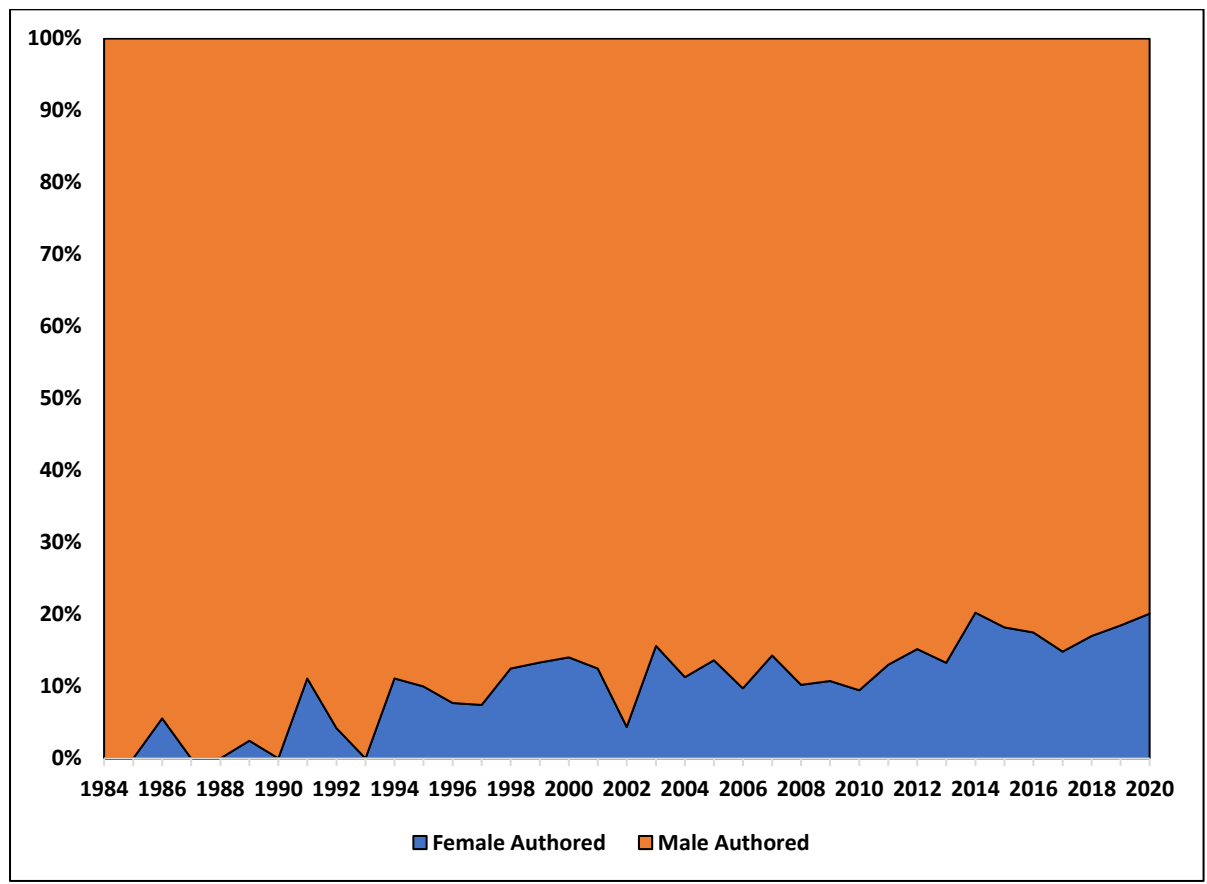

**Fig.6: Gender distribution of research publications in FGCS**

*3.9 Social Media visibility of research publications*

During the recent years, altmetric has emerged as an alternative measure of impact of research publications [39], [40], [41]. It tries to measure the activity around research articles in social media and academic social networks etc. [42]. Altmetrics has gained popularity more during the last decade. We have tried to measure the altmetric activity around research publications in FGCS by obtaining the data from popular altmetric aggregator, Altmetric.com [43], [44]. Since the altmetric is a recent phenomenon, we have tried to measure the altmetric activity only for the recent period of 2011 to 2020. The publication DOI is passed on the Altmetric.com, which returns the tracked altmetric data from different platforms, such as Twitter, Facebook, News Mediums, Blog Platforms, Wikipedia and Mendeley etc. Table 5 presents the coverage percentage and average mentions per paper in different platforms for the FGCS publications.

It was found that out of 3,862 publication records in FGCS during 2011-20, a total of 1,274 are found covered by Altmetric.com, which is 32.98% coverage. **Table 6** presents the coverage and average mentions per paper for FGCS articles in different platforms. Mendeley has highest coverage (32.91%) and average mentions per paper value (91.69) among all platforms. Twitter platform has coverage of 550 articles (14,24% coverage) with average tweets per paper of 4.18. Facebook, News and Blog platforms have relatively lower values. The coverage and average mentions per paper values are, however, similar to those observed in studies for many other reputed journals in Computer Science.

Table 6: Altmetric Coverage of publications in FGCS

| Platform | Coverage and Average mentions | |
|---|---|---|
| **Twitter** | Articles covered in Twitter | 550 |
| | Coverage (%) | 14.24% |
| | Avg. Mentions/ paper | 4.18 |
| **Facebook (FB)** | Articles covered in FB | 41 |
| | Coverage (%) | 1.06% |
| | Avg. Mentions/ paper | 1.12 |
| **News** | Articles covered in News | 35 |
| | Coverage (%) | 0.91% |
| | Avg. Mentions/ paper | 2.31 |
| **Blog** | Articles covered in Blog | 18 |
| | Coverage (%) | 0.47% |
| | Avg. Mentions/ paper | 1.33 |
| **Wikipedia** | Articles covered in Wikipedia | 34 |
| | Coverage (%) | 0.88% |
| | Avg. Mentions/ paper | 1.23 |
| **Mendeley** | Articles covered in Mendeley | 1271 |
| | Coverage (%) | 32.91% |
| | Avg. Mentions/ paper | 91.69 |

Note: Out of 3,862 publications during 2011-20, a total of 1,274 publications are found covered by Altmetric.com, which is 32.98% coverage.

*3.10 SDG Connection of the publications*

The Sustainable Development Goals (SDGs), also referred to as the Global Goals, were adopted by the United Nations in 2015. There are a total of 17 goals along with 169 indicators which are targeted to be achieved by 2030. These goals are focussed on overall human well-being (they address issues like affordable and clean energy, sustainable cities and communities, climate action, good health and well-being etc.). Achieving these goals require research and development and hence contribution of researchers across the world is very important [45]. It has been reported that the research publications related to SDGs have been increasing widely [46]. Studies like [47] and [48] have analysed the technology-oriented publications related to SDGs. FGCS is an important future oriented journal and hence is expected to be publishing research and technological advancements in the area of SDGs. Motivated by this, we have tried to identify what number of papers in FGCS are related to some SDG. Identifying whether a publication is related to a SDG is a challenge. Fortunately, the Dimensions database provides an automated classification of the articles in different SDG(s) based on the content of the research articles. Therefore, we used this information to map which publications from FGCS are related to one or more SDGs. **Figure 7** presents a tree map of the publications from FGCS classified under different SDGs. It can be observed that the Goal 7 (Affordable and Clean Energy), Goal 11 (Sustainable Cities and Communities) and Goal 3 (Good health and well-being) have maximum number of publications classified under them. These areas have 355, 64 and 54 papers, respectively. The results indicate that FGCS publishes a good amount of research publications on research and technological advancements in Affordable and Clean Energy, Sustainable Cities and Communities and Good health and Well-being. This is an indication of future alignment and forward-looking scope and agenda of the FGCS journal.

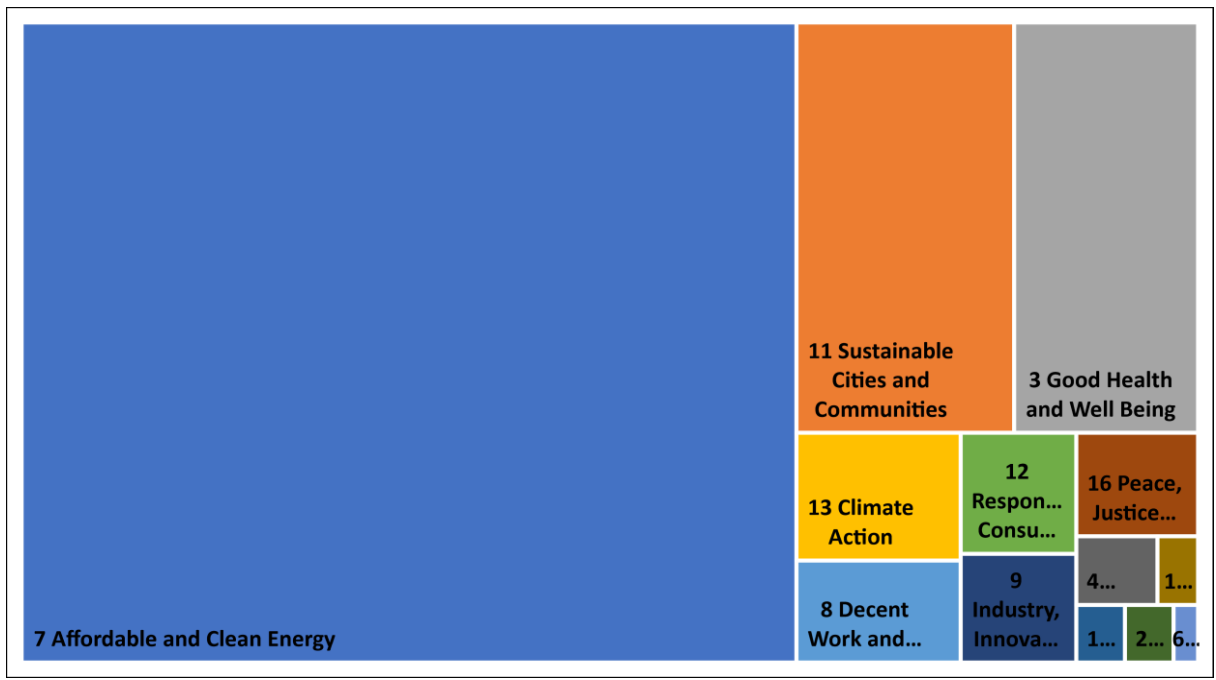

Fig. 7: Publications related to SDGs in FGCS during 2015-2020

## 4. Text-integrated Path Analysis

*4.1 Evolution of Concepts: Insights from text-integrated Path Analysis*

The first step is creation of citation network of publications in FGCS. This network is found to be consisting of 4095 papers and 9233 links. It is shown **in Figure 8.**

Here, colors of nodes represent different clusters. However, we are not attempting to discuss the cluster analysis in this work as our major focus is evolutionary trajectory analysis using concepts.

As mentioned in methodology, the different paths obtained are SPC FW, SPC BW, SPC KR, SPC CPM, FV FW, FV BW, FV KR, FV CPM where FW indicates forward search, BW indicates backward search, KR indicate key-route search, CPM indicates and critical path method. In this work, as key-route (local) and key-route (global) provided same results, we used KR to represent both. Subsequently, **Figure 9** shows SPC FW, SPC BW, SPC KR, SPC CPM paths while **Figure 10** shows FV FW, FV BW, FV KR, FV CPM paths.

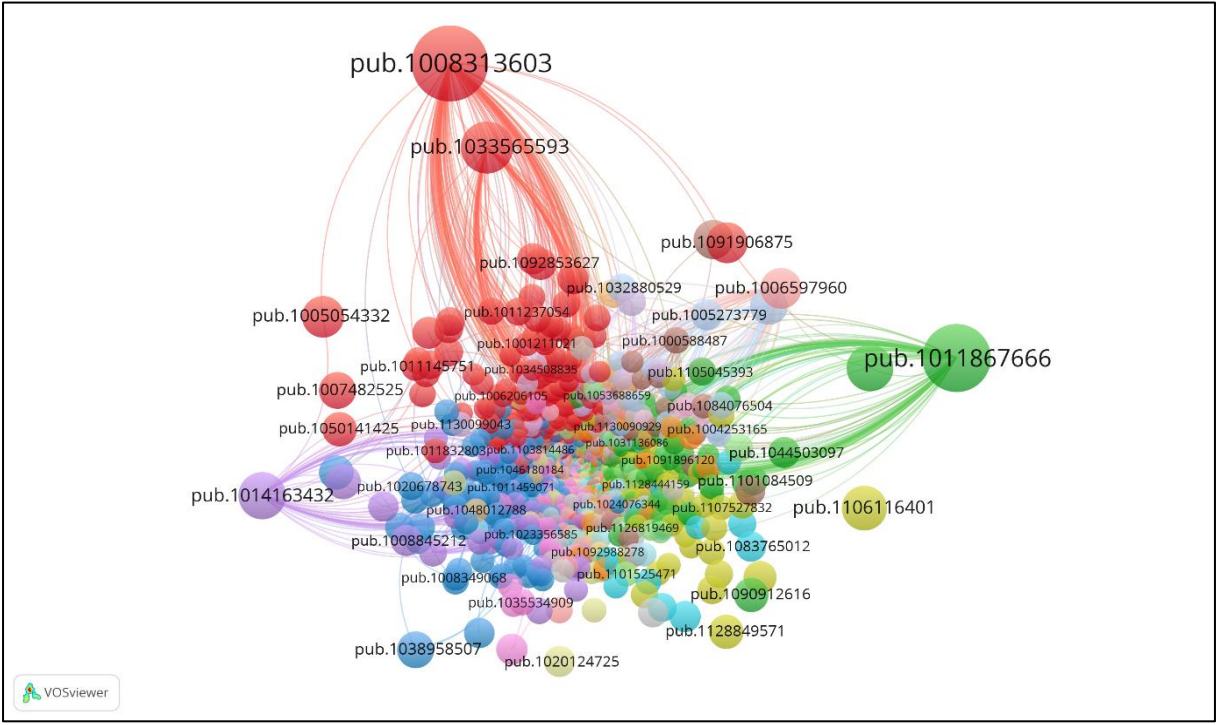

**Fig. 8: Citation network of FGCS published works during 1984-2020**

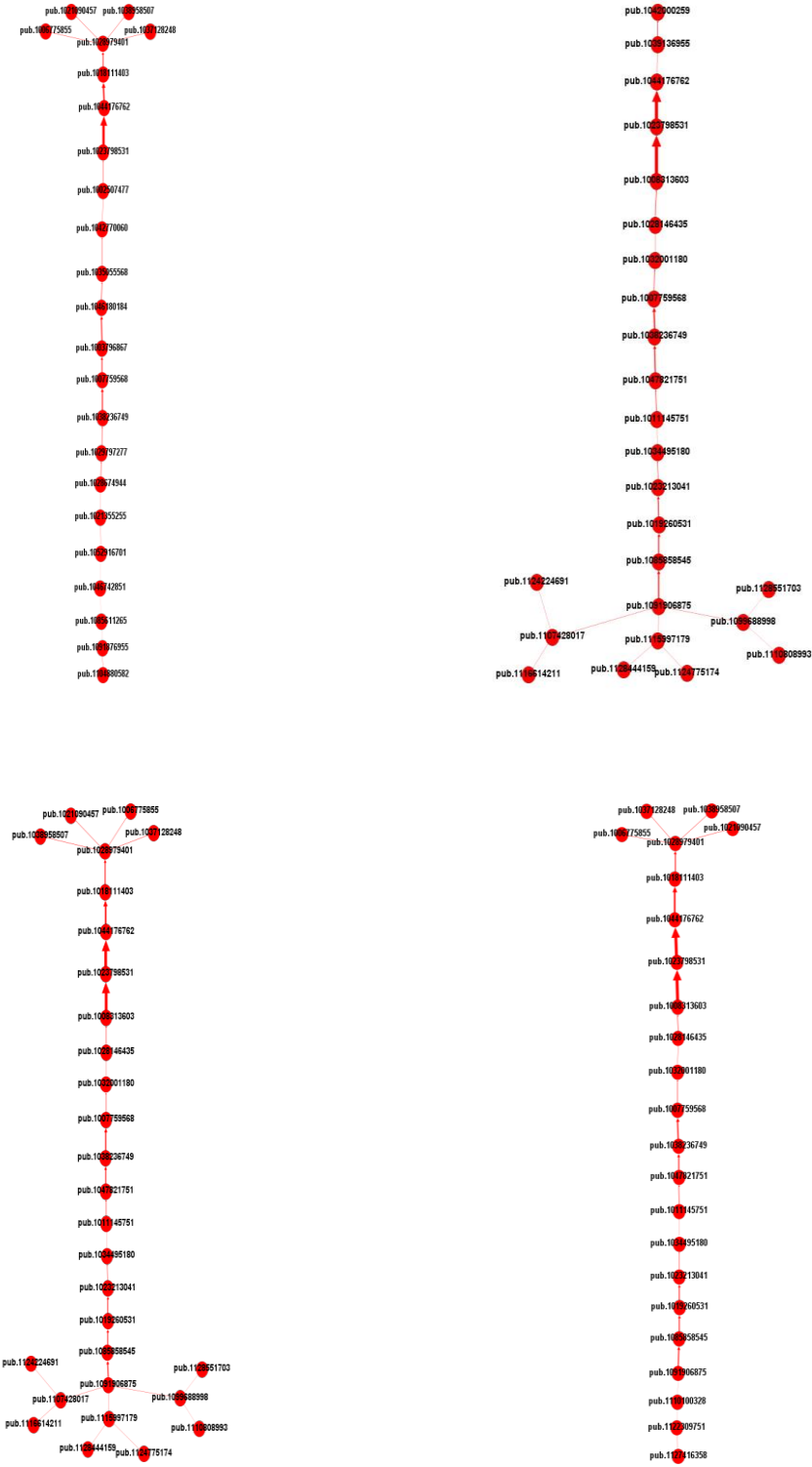

Fig. 9: SPC Forward path (top left), SPC Backward path (top right), SPC Key-route path (bottom left) and SPC Critical path (bottom right)

**Fig. 10: FV Forward path (top left), FV Backward path (top right), FV Key-route path (bottom left) and FV Critical path (bottom right)**

As mentioned in the procedure, before going for the creation of concept evolution paths, we have to select the paths that are having adequate uniqueness from others to avoid redundancy in analysis. For that uniqueness, indices with respect to every pair of paths is created and thus we get U-index matrix (**Table 7**), which is shown below:

**Table 7: U-index matrix for paths in FGCS**

|  | SPC FW | SPC BW | SPC KR | SPC CPM | FV FW | FV BW | FV KR | FV CPM |
|---|---|---|---|---|---|---|---|---|
| **SPC FW** | * | 0.917 | 0.808 | 0.804 | 0.97 | 1 | 1 | 0.768 |
| **SPC BW** | * | * | **0.574** | 0.708 | 1 | 0.971 | 1 | 0.841 |
| **SPC KR** | * | * | * | **0.615** | 1 | 0.949 | 1 | 0.813 |
| **SPC CPM** | * | * | * | * | 1 | 0.939 | 1 | 0.754 |
| **FV FW** | * | * | * | * | * | 0.95 | 1 | 1 |
| **FV BW** | * | * | * | * | * | * | 1 | 1 |
| **FV KR** | * | * | * | * | * | * | * | 0.984 |
| **FV CPM** | * | * | * | * | * | * | * | * |

The values indicate that apart for the pairs SPC BW- SPC KR and SPC KR- SPC CPM, all other values are satisfactorily high enough to indicate distinction. As U value of SPC-KR and SPC-BW is close to 0.5, on inspection it is found that most of the works in SPC-BW is found in SPC-KR and as SPC-KR has more works, so we can drop SPC BW from our analysis as analysis of SPC KR might include most of the evolutionary information of SPC BW too. Similar is the case of SPC KR and SPC CPM. Thus, inclusion of SPC KR in analysis might cover the evolutionary information of both SPC BW and SPC CPM. Now we can proceed for the creation of concept evolution paths and its analysis.

*4.2 Concept evolution path with respect to SPC FW*

From **Figure 11**, we can see the evolution of concepts. In this concept evolution path, 'cloud computing' occupies a special place. Firstly, a knowledge convergence from some other fields such as management and electrical systems that might be treated as application areas of computer systems towards cloud computing is visible. Among the application fields, power grid system is a networked system. When it comes to computing, knowledge flow is witnessed from grid computing (in its earlier years) to cloud computing is also found as a part of the initial convergence. Then, Knowledge flow from cloud computing to advanced cloud computing systems and to grid computing systems and then a recombination with grid computing is visible. The cycle that starts from cloud computing indicates that knowledge advancement propagated from cloud computing led to the enhancement of cloud computing itself. After that knowledge flow extends towards application areas that again deal with 'network paradigm' such as scheduling in 'power grids', heterogenous distributed systems and big data platform.

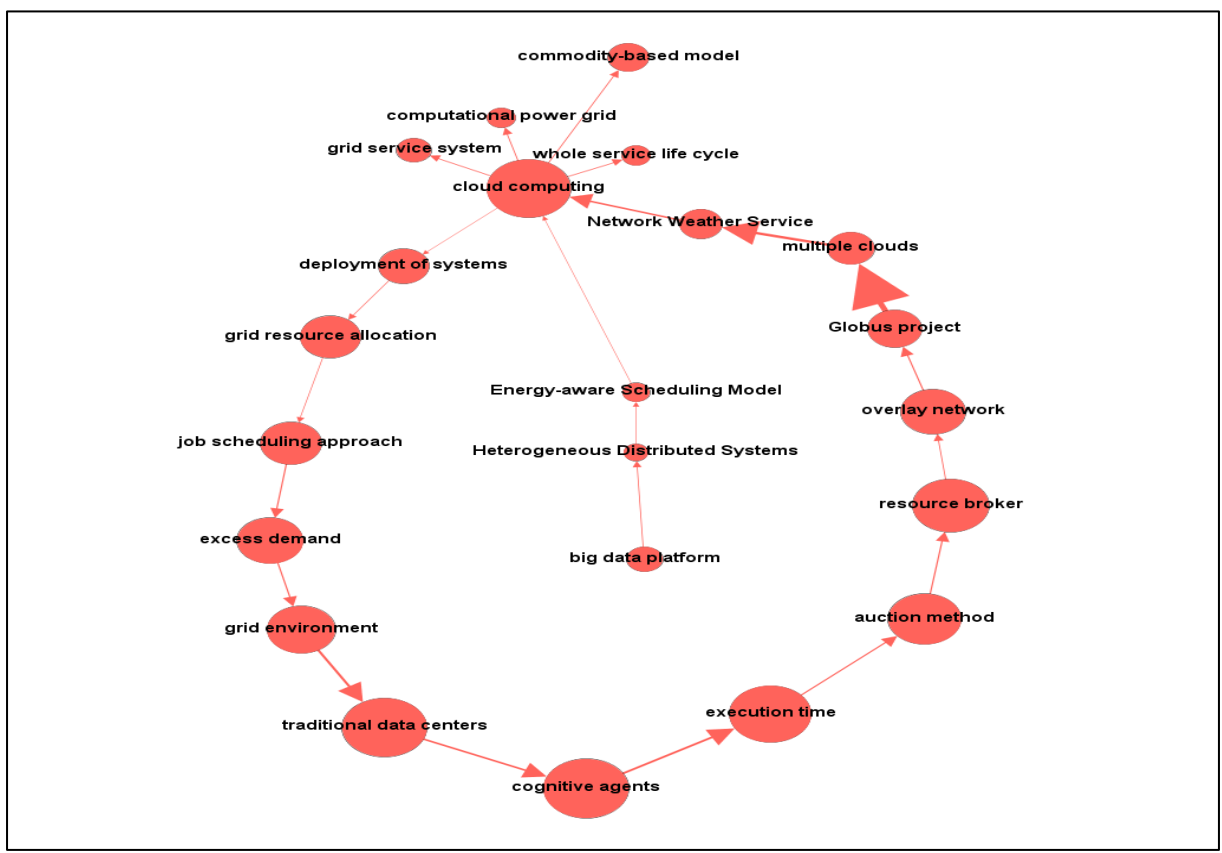

**Fig. 11: Concept evolution path with respect to SPC FW**

*4.3 Concept evolution path with respect to SPC KR*

In fig. 12, though no concepts are exactly phrased as 'cloud computing' in the concept evolution path associated with SPC KR path, most of the concepts are related to cloud computing. Knowledge flow between cloud technologies and some application areas such as management, power grid and other networking aspects are profound in this concept evolution path. Knowledge exchange between advanced cloud computing systems and grid computing systems is found in this path also just like concept evolution path with respect to SPC FW. Network management applications related to scheduling and resource allocation is also found in this path. Towards the end of this path, some important divergences to IoT services, network security and management, virtual machines and business processes are observed. These are the application areas in which major FGCS technologies especially the ones related to 'networking paradigm' can make a huge impact.

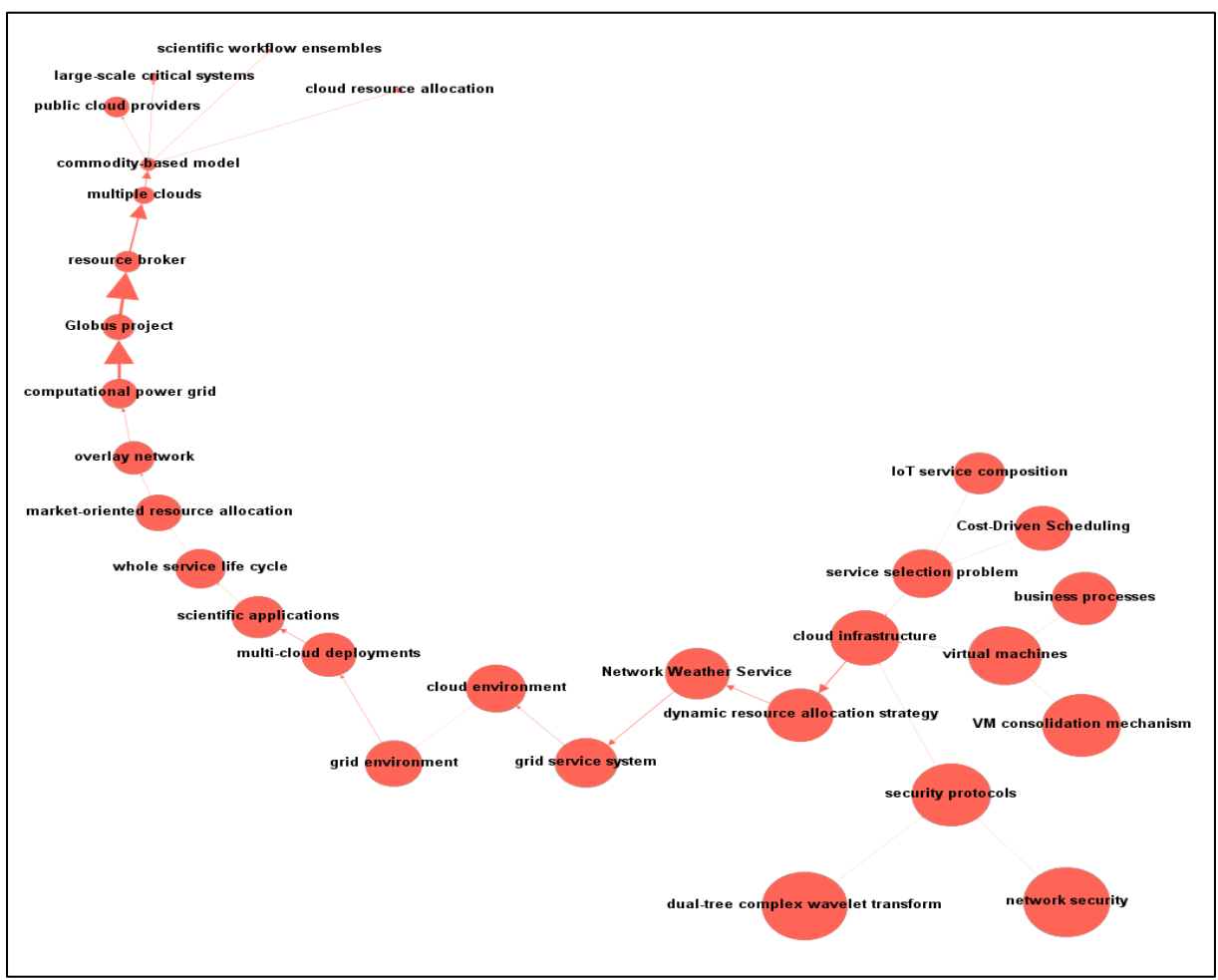

**Fig. 12: Concept evolution path with respect to SPC KR**

*4.4 Concept evolution path with respect to FV FW*

Advancement of cloud computing systems with the help of knowledge flow between grid computing and cloud computing is witnessed in the case of FV FW path too. This evolution is characterized by the developments in grid management, architectural changes for data access and management. These developments led to 'mobile cloud computing', which is a combination of mobile computing and cloud computing for bringing rich computational resources to mobile users, network operators, as well as cloud computing providers. This is a perfect example of technological cross over, which is disseminated by the FGCS journal. The evolution is recently heading towards Open Grid Service Architecture Data Access and Integration (OGSA-DAI) technology that aims to providing interfaces to heterogeneous data sources on Grids for integration them. Profound evolution of computing systems towards 'network paradigm' is evident from this path too and the well-deserved emphasis on security is also observable from the presence of the concept 'new encryption mechanism'.

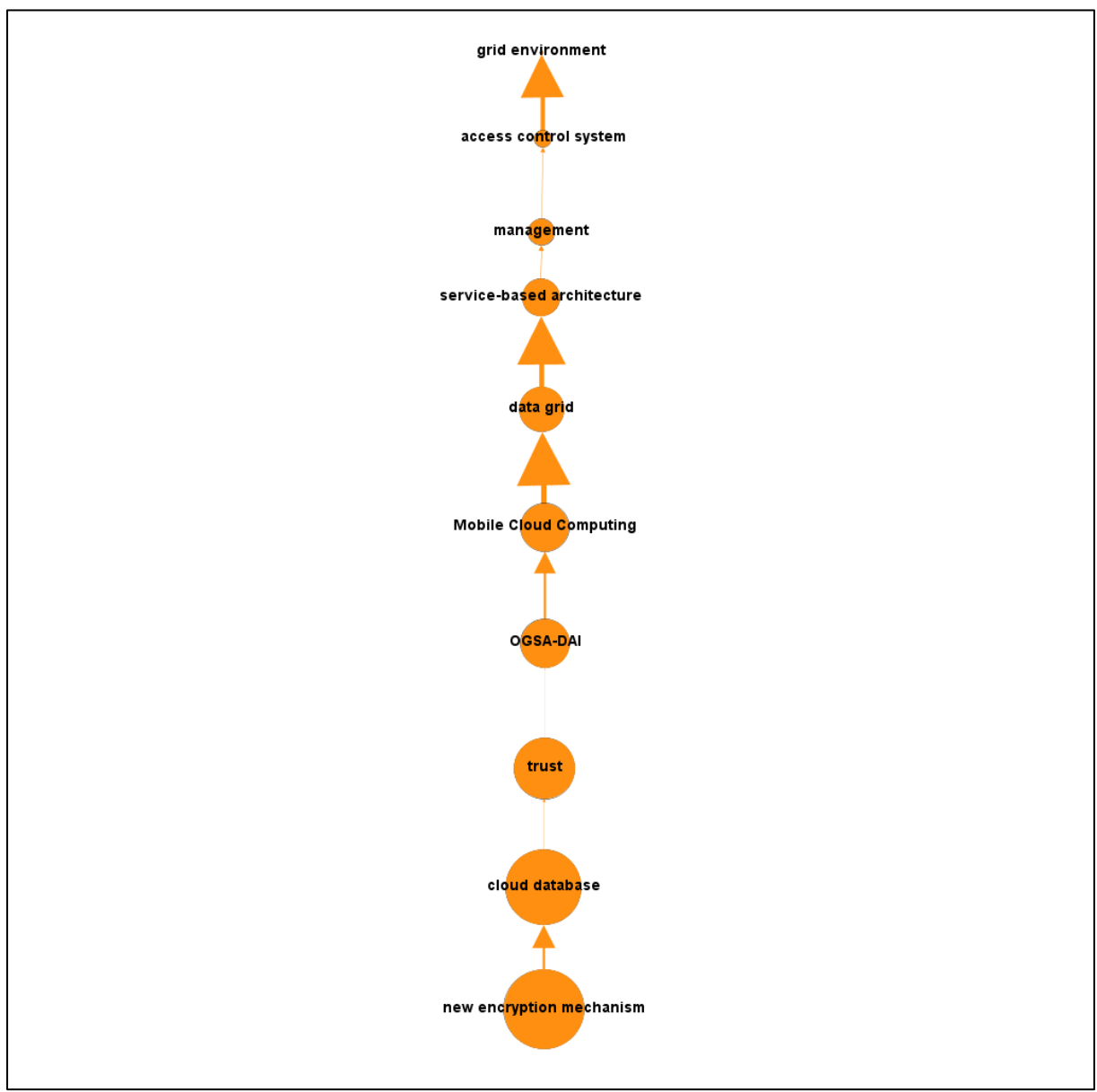

**Fig. 13: Concept evolution path with respect to FV FW**

*4.5 Concept evolution path with respect to FV BW*

This path is mainly related to the evolution of IoT systems. At first a knowledge flow from management to IoT, especially in terms of network techniques-based resource allocation and management is witnessed. Knowledge exchange between IoT and Cloud computing is visible in this concept evolution path. As seen in **Figure 14**, a self-loop in the term 'Internet of Things' is due to the citation of a paper with 'Internet of Things' as top relevant concept by another paper with the same phrase as top relevant concept. As IoT systems get more enhanced with advanced cloud computing technologies, some interesting divergences that are useful for 'networked systems' is visible. These are network security (indicated by the concept 'new encryption mechanism'), blockchain technology, etc. This evolution path highlights the importance of IoT technologies (aided with advanced computing technologies) to revolutionize many application areas including business management and banking.

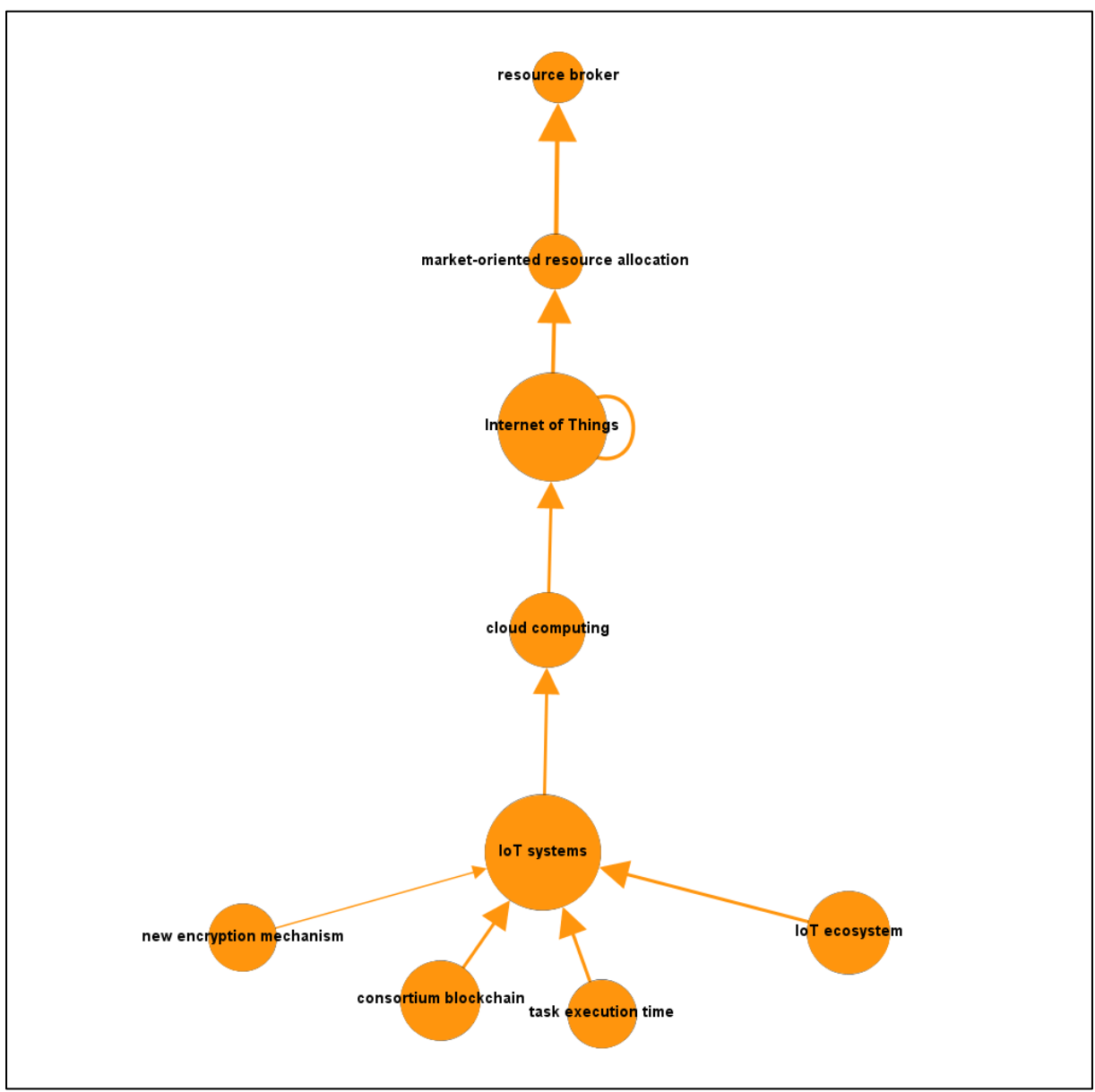

**Fig. 14: Concept evolution path with respect to FV BW**

*4.6 Concept evolution path with respect to FV KR*

The evolution path shown in **Figure 15** is also related to IoT. Knowledge flow from cloud computing is indicated by the presence of concept 'integration of cloud'. Knowledge exchange from 'data mining' and resource management and its role in evolution of IoT is clear from this evolution path. Importance of networking aspect in IoT is so profound that major developments in it have an impact in distributed computing itself. Recent evolution in computing is found to be related to 'Edge computing' technology, a network-based distributed computing technology. This is another evidence on how well the journal of FGCS is capturing important developments in the themes that comes under its scope.

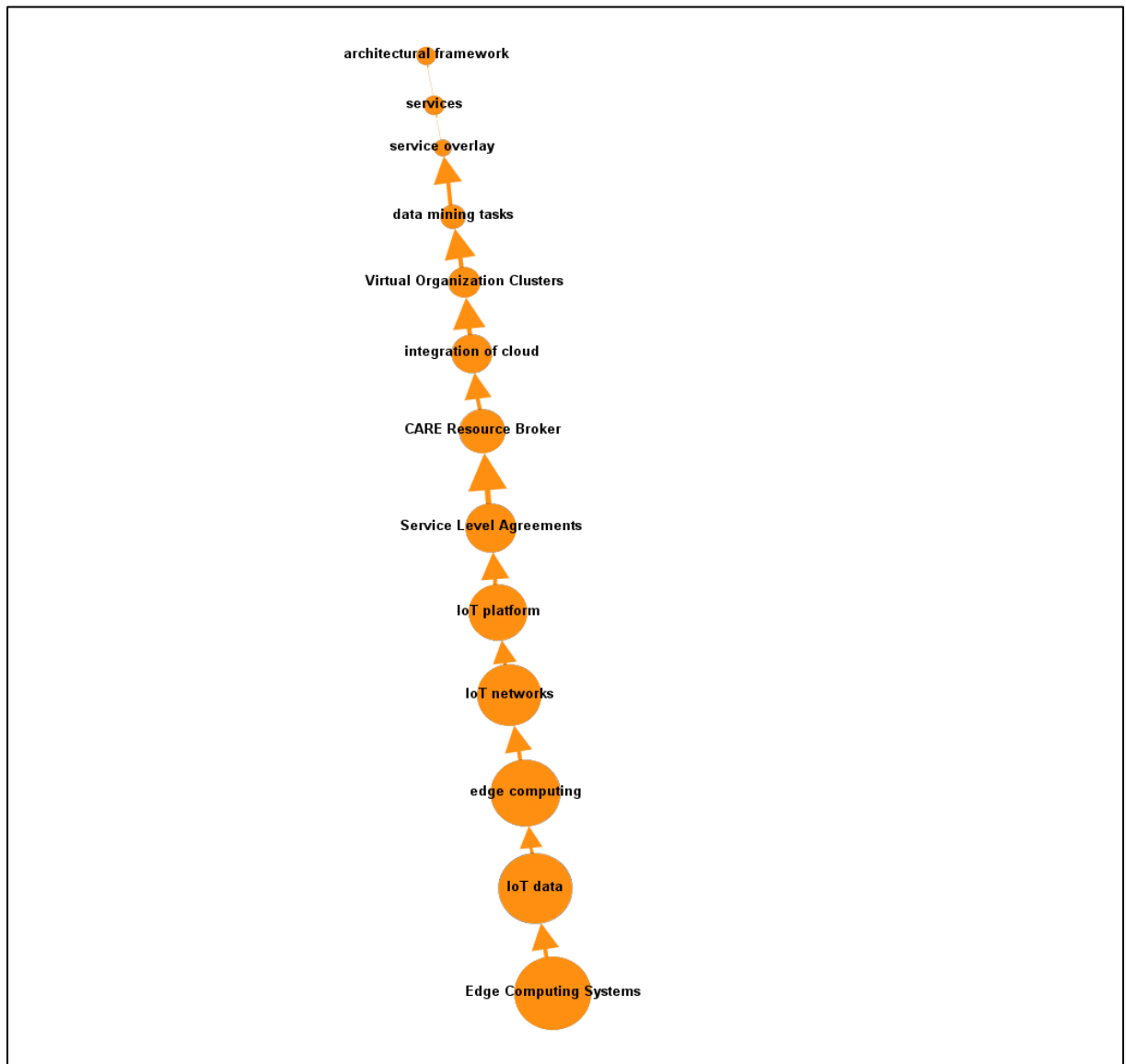

**Fig. 15: Concept evolution path with respect to FV KR**

*4.7 Concept evolution path with respect to FV CPM*

The path shown in **Figure 16** gives more information about the knowledge flow to and from 'cloud computing' to various applications and technologies. The impact of developments in other fields such as virtual machines on cloud computing and vice versa is also observed. It also shows the extension chain of knowledge to big data and data mining applications. Applications of cloud computing and virtual reality paradigm in management applications is also evident from this somewhat more comprehensive concept evolution path. Especially, the divergence from virtual machines towards different service level applications of cloud computing is very interesting as well as useful for both academic and industrial practitioners. Knowledge extension from cloud computing to multi-domain networks is witnessed and a divergence towards 'service function chains', 'distributed machine learning' and 'edge computing systems' is visible. So, this concept evolution path captured the important developments in 'network paradigm', 'virtual reality' and major futuristic computing technologies in revolutionizing the current state-of-the-art in their respective fields and also the application areas.

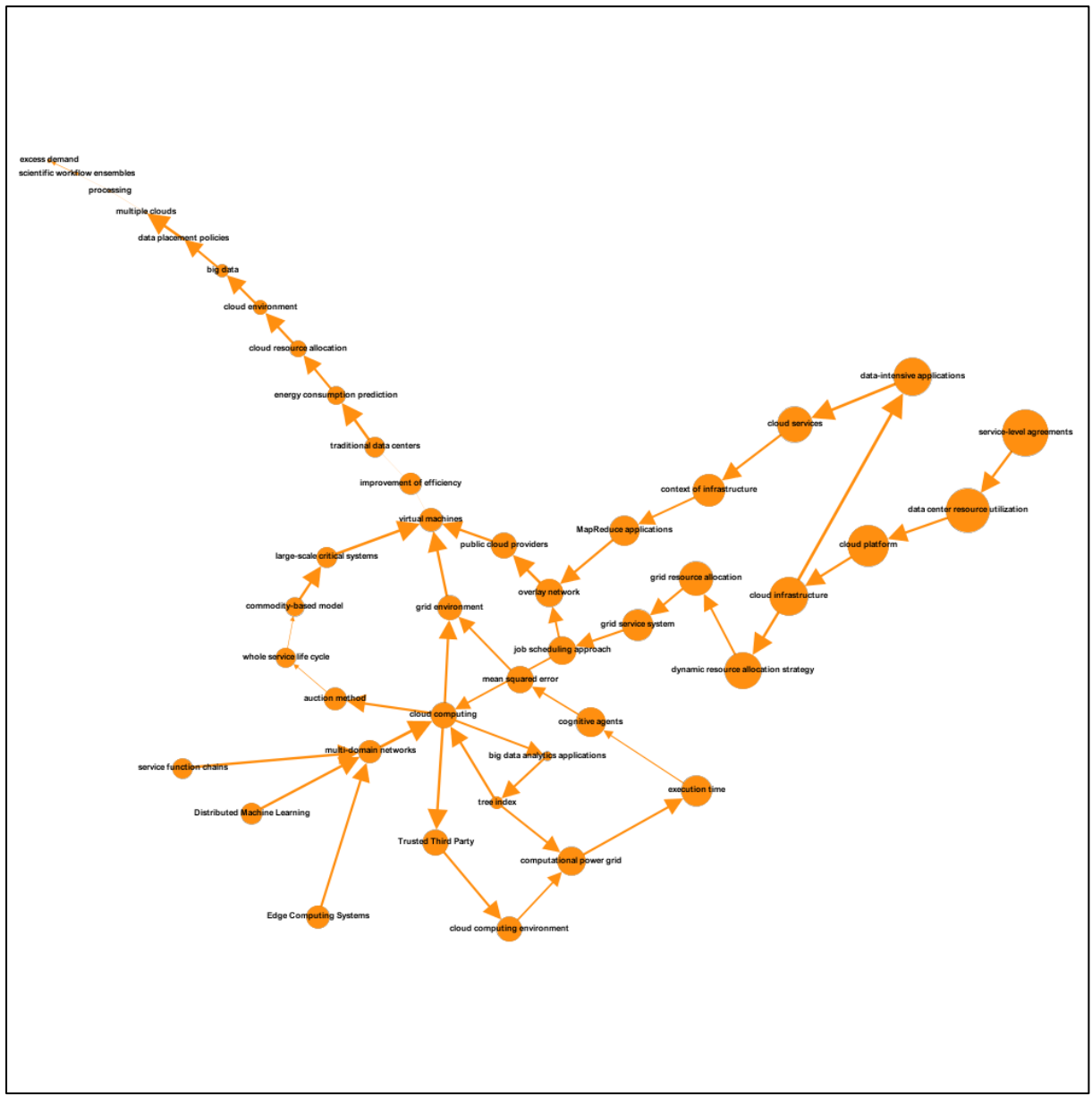

**Fig. 16: Concept evolution path with respect to FV CPM**

From all the concept evolution paths, a general observation is that cloud computing and also grid computing plays a pivotal role in many evolutionary trajectories, and it is very much vital for FGCS. Technologies associated with 'Network paradigm', especially 'distributed computing' such as edge computing is found to be highly promising one that could revolutionize future computer systems as well as systems of generic appeal like IoT and business management systems. It will be an interesting endeavor to dig deeper into these areas specifically to capture the specific developments that can work wonders.

**Conclusion**

The article presented analytical results of publication and citation patterns, authorship structure, collaboration patterns, major funding sources, gender distribution, social- media visibility and UN SDG connections of papers published in FGCS etc. In addition, it traced the major concept evolution trajectories through application of a text-integrated path analysis method. It is found that number of publications in FGCS grew by a factor of 45 from 1984 to 2020, with phenomenal growth observed in the 1988-89, 1991-92 and 2017-18. Citations to the journal also grew rapidly and the citations per paper has also improved significantly. The FGCS now has an h-index of 142. China, United States and United Kingdom are found to be the major contributors to FGCS publications. A clear pattern of shift from single authored to multi-authored papers is observed during 1984-2020. Similarly, papers from single institution have decreased while multi-institutional and internationally collaborated papers have increased significantly. The open access availability of papers in FGCS has also increased over time, reaching to about 20-30% in the recent time. In terms of gender distribution, the proportion of female first authored papers have also increased to reach to about 20% as in 2020. A good number of FGCS publications are found to be directly connected with Sustainable Development Goals. Cloud, Grid, Big Data, IoT, Security, Distributed Systems are found to be the major conceptual themes discussed in FGCS publications. Knowledge flows from 'cloud' to 'grid' and 'big data' are visible among the concept evolution trajectories. Similarly, knowledge extension from 'cloud computing' to 'multi-domain networks' is witnessed and a divergence towards 'service function chains', 'distributed machine learning' and 'edge computing systems' is visible. Overall, cloud computing is found to play a pivotal role in many concept evolution trajectories. The article thus presents a comprehensive bibliometric analysis and text-integrated path analysis of concept evolution trajectories of FGCS, which uncovers the publication patterns and conceptual structure of FGCS.

**References**

[1]   Christian Herzog, Daniel Hook, Stacy Konkiel, Dimensions: Bringing down barriers between scientometricians and data, Quantitative Science Studies, 1(1) (2020) 387-395.

[2]   Vivek Kumar Singh, Mousumi Karmakar, Jacqueline Leta, Phillip Mayrr, The Journal Coverage of Web of Science, Scopus and Dimensions: A Comparative Analysis, Scientometrics, 126(6) (2021) 5113-5142.

[3]   Naveen Donthu, Satish Kumar, Nitesh Pandey, Prashant Gupta, Forty years of the International Journal of Information Management: A bibliometric analysis, Journal of Information Management, 57 (2021) 102307 10.1016/j.ijinfomgt.2020.102307

[4]   Jose M. Merigo, Witold Pedrycz, Richard Weber, Catalina de la Sotta, Fifty years of Information Sciences: A bibliometric overview, Information Sciences, 432 (2018) 245-268

[5] Raghu Raman, Prashasti Singh, Vivek Kumar Singh, Ricardo Vinuesa, Prema Nedungadi, Understanding the Bibliometric Patterns of Publications in IEEE Access, IEEE Access, 10 (2022) 35561-35577.

[6] Shiwangi Singh, Sanjay Dhir, V. Mukunda Das, Anuj Sharma, Bibliometric overview of the Technological Forecasting and Social Change journal: Analysis from 1970 to 2018, Technological Forecasting and Social Change, 154 (2020) 119963 10.1016/j.techfore.2020.119963

[7] Derek J. De Solla Price, Networks of scientific papers: The pattern of bibliographic references indicates the nature of the scientific research front, Science, 149(3683) (1965) 510-515

[8] Eugene Garfield, Irving H. Sher, Richard J. Torpie, The use of citation data in writing the history of science, Institute for Scientific Information Inc Philadelphia PA, 1964

[9] Norman P. Hummon, Patrick Dereian, Connectivity in a citation network: The development of DNA theory, Social Networks, 11(1) (1989) 39-63.

[10] Andrea Mina, Ronald Ramlogan, Gindo Tampubolon, J. Stanley Metcalfe, Mapping evolutionary trajectories: Applications to the growth and transformation of medical knowledge, Research policy, 36(5) (2007) 789-806.

[11] Gindo Tampubolon, Ronald Ramlogan, Networks and temporality in the development of a radical medical treatment, Graduate Journal of Social Science, 4(1) (2007) 54-77.

[12] Sung Jun Jo, Chang-Wook Jeung, Sunyoung Park, Hea Jun Yoon, Who is citing whom: Citation network analysis among HRD publications from 1990 to 2007, Human resource development quarterly, 20(4) (2009) 503-537.

[13] Tom Brughmans, Networks of networks: A citation network analysis of the adoption, use, and adaptation of formal network techniques in archaeology, Literary and Linguistic Computing, 28(4) (2013) 538-562.

[14] Vladimir Batagelj, Efficient algorithms for citation network analysis, 2003, arXiv preprint, arXiv:cs/0309023.

[15] Hiran H. Lathabai, Susan George, Thara Prabhakaran, Manoj Changat, An integrated approach to path analysis for weighted citation networks, Scientometrics, 117(3) (2018) 1871-1904.

[16] John S. Liu, Louis YY Lu, An integrated approach for main path analysis: Development of the Hirsch index as an example, Journal of the American Society for Information Science and Technology, 63(3) (2012) 528-542.

[17] Thara Prabhakaran, Hiran H. Lathabai, Manoj Changat, Detection of paradigm shifts and emerging fields using scientific network: A case study of Information Technology for Engineering, Technological Forecasting and Social Change, 91 (2015) 124-145.

[18] Stephen P. Borgatti, Centrality and network flow, Social networks, 27(1) (2005) 55-71.

[19] Giovanni Scardoni, Carlo Laudanna, Centralities based analysis of complex networks, New frontiers in graph theory (2012) 323-348.

[20] Hiran H. Lathabai, Thara Prabhakaran, Manoj Changat, Centrality and flow vergence gradient based path analysis of scientific literature: A case study of biotechnology for engineering, Physica A: Statistical Mechanics and its Applications. 429 (2015) 157-168.

[21] Hiran H. Lathabai, Thara Prabhakaran, Manoj Changat, Contextual productivity assessment of authors and journals: a network scientometric approach. Scientometrics, 110(2) (2017) 711-737.

[22] Thara Prabhakaran, Hiran H. Lathabai, Susan George, Manoj Changat, Towards prediction of paradigm shifts from scientific literature, Scientometrics, 117(3) (2018) 1611-1644.

[23] Elizabeth S. Vieira, Jose ANF Gomes, Citations to scientific articles: Its distribution and dependence on the article features, Journal of Informetrics 4(1) (2010) 1-13.

[24] Heidi Ledford, Team science, Nature, 525(7569) (2015) 308.

[25] Kara L. Hall, Amanda L. Vogel, Grace C. Huang, Katrina J. Serrano, Elise L. Rice, Sophia P. Tsakraklides, Stepehen M. Fiore, The science of team science: A review of the empirical evidence and research gaps on collaboration in science. American psychologist, 73(4) (2018) 532.

[26] Michael Gordon, A critical reassessment of inferred relations between multiple authorship, scientific collaboration, the production of papers and their acceptance for publication, Scientometrics, 2(3) (1980) 193-201.

[27] John Smart, Alan Bayer, Author collaboration and impact: A note on citation rates of single and multiple authored articles, Scientometrics, 10(5-6) (1986) 297-305.

[28] Wolfgang Glänzel, National characteristics in international scientific co-authorship relations, Scientometrics, 51(1) (2001) 69-115.

[29] Peter Van Deb Besselaar, Sven Hemlin, Inge Van Der Weijden, Collaboration and competition in research, Higher Education Policy, 25(3) (2012) 263-266.

[30] Sunny Li Sun, Mike W. Peng, Ruby P. Lee, Weiqiang Tan, Institutional open access at home and outward internationalization, Journal of World Business, 50(1) (2015) 234-246.

[31] Mikael Laakso, Patrik Welling, Helena Bukvova, Linus Nyman, Bo-Christer Björk, Turid Hedlund, The development of open access journal publishing from 1993 to 2009, PloS One, 6(6) (2011) e20961.

[32] Leslie Chan, Barbara Kirsop, Sely Maria de Souza Costa, Subbiah Arunachalam, Improving access to research literature in developing countries: challenges and opportunities provided by Open Access (2005).

[33] Heather Piwowar, Jason Priem, Vincent Larivière, Juan Pablo Alperin, Lisa Matthias, Bree Norlander, Ashley Farley, Jevin West, Stefanie Haustein, The state of OA: a large-scale analysis of the prevalence and impact of Open Access articles, PeerJ, 6 (2018) e4375.

[34] Vivek Kumar Singh, Rajesh Piryani, Satya Swarup Srichandan, The case of significant variations in gold–green and black open access: evidence from Indian research output, Scientometrics, 124(1) (2020) 515-531

[35] Satya Swarup Srichandan, Rajesh Piryani, Vivek Kumar Singh, Sujit Bhattacharya, The Status and Patterns of open Access in Research Output of Most Productive Indian Institutions, Journal of Scientometric Research, 9(2) (2020) 96-110.

[36] Jyoti Paswan, Vivek Kumar Singh, Gender and research publishing analyzed through the lenses of discipline, institution types, impact and international collaboration: a case study from India. Scientometrics, 123(1) (2020) 497-515.

[37] Mike Thelwall, Carol Bailey, Meiko Makita, Pardeep Sud, Devika P. Madalli, Gender and research publishing in India: Uniformly high inequality?, Journal of informetrics, 13(1) (2019a) 118-131.

[38] Mike Thelwall, Carol Bailey, Catherine Tobin, Noel-Ann Bradshaw, Gender differences in research areas, methods and topics: Can people and thing orientations explain the results?, Journal of Informetrics, 13(1) (2019)149-169.

[39] Heather Piwowar, Introduction altmetrics: What, why and where? Bulletin of the American Society for Information Science and Technology, 39(4) (2013) 8-9.

[40] Lutz Bornamann, Do altmetrics point to the broader impact of research? An overview of benefits and disadvantages of altmetrics, Journal of Informetrics, 8(4) (2014) 895-903.

[41] Jason Priem, Paul Groth, and Dario Taraborelli, The altmetrics collection, PloS one 7(11) (2012) e48753.

[42] Cassidy R. Sugimoto, Sam Work, Vincent Larivière, and Stefanie Haustein, Scholarly use of social media and altmetrics: A review of the literature, Journal of the association for information science and technology, 68(9) (2017) 2037-2062.

[43] José Luis Ortega, Reliability and accuracy of altmetric providers: a comparison among Altmetric. com, PlumX and Crossref Event Data, Scientometrics 116(3) (2018) 2123-2138.

[44] Mousumi Karmakar, Sumit Kumar Banshal, and Vivek Kumar Singh, A large-scale comparison of coverage and mentions captured by the two altmetric aggregators: Altmetric. com and PlumX, Scientometrics 126(5) (2021) 4465-4489.

[45] Caroline S. Armitage, Marta Lorenz, and Susanne Mikki, Mapping scholarly publications related to the Sustainable Development Goals: Do independent bibliometric approaches get the same results?, Quantitative Science Studies 1(3) (2020) 1092-1108.

[46] Christine Mischede, The sustainable development goals in scientific literature: A bibliometric overview at the meta-level, Sustainability 12(11) (2020) 4461.

[47] Ricardo Vinuesa, Hossein Azizpour, Iolanda Leite, Madeline Balaam, Virginia Dignum, Sami Domisch, Anna Felländer, Simone Daniela Langhans, Max Tegmark, Francesco Fuso Nerini, The role of artificial intelligence in achieving the Sustainable Development Goals, Nature communications 11(1) (2020) 1-10.

[48] Shivam Gupta, Simone D. Langhans, Sami Domisch, Francesco Fuso-Nerini, Anna Felländer, Manuela Battaglini, Max Tegmark, and Ricardo Vinuesa, Assessing whether artificial intelligence is an enabler or an inhibitor of sustainability at indicator level, Transportation Engineering 4 (2021) 100064.